\documentclass[twocolumn, Journal]{IEEEtran}
\IEEEoverridecommandlockouts
\usepackage{float}
\usepackage{bm}
\usepackage{subfig}
\usepackage{color}
\usepackage{graphicx}
\usepackage{amsmath}
\usepackage{amssymb}
\usepackage{algorithm}
\usepackage{algorithmic}
\usepackage{amsmath}
\usepackage{multirow}
\usepackage{booktabs}
\usepackage{array}
\usepackage{amsthm}
\usepackage{stfloats}
\usepackage{caption}
\usepackage{cuted}
\usepackage{setspace}
\usepackage{flushend}

\usepackage{lastpage}

\newcommand{\be}{\begin{equation}}
\newcommand{\ee}{\end{equation}}
\newcommand{\bea}{\begin{eqnarray}}
\newcommand{\eea}{\end{eqnarray}}
\newcommand{\ba}{\begin{array}}
\newcommand{\ea}{\end{array}}

\newcommand\ddfrac[2]{\frac{\displaystyle #1}{\displaystyle #2}}

\title{Joint Waveform and Beamforming Design in RIS-ISAC Systems: A Model-Driven Learning Approach
\thanks{Part of this paper was presented at the IEEE Vehicular Technology Conference (VTC2024-Fall) \cite{VTC2024}.}
\thanks{Manuscript received September 26, 2024; revised December 22, 2024; accepted January 20, 2025. This work is supported in part by the National Natural Science Foundation of China (Grant No. 62371090, 62471086, and 62271163), in part by Fundamental Research Funds for the Central Universities (Grant No. DUT24ZD125 and DUT24RC(3)005), and in part by Liaoning Applied Basic Research Program (2023JH2/101700364). The work of Rang Liu is supported by the U.S. National Science Foundation under Grant CCF-2225575 and Grant CCF-2322191.
The associate editor coordinating the review of this article and approving it for publication was Dr. Fan Liu. \textit{(Corresponding author: Ming Li.)}}
\thanks{P. Jiang, M. Li, and W. Wang are with the School of Information and Communication Engineering, Dalian University of Technology, Dalian 116024, China (e-mail: pengjiang@mail.dlut.edu.cn; mli@dlut.edu.cn; wangwei2023@dlut.edu.cn).}
\thanks{R. Liu was with the School of Information and Communication Engineering, Dalian University of Technology, Dalian 116024, China. She is currently with the Center for Pervasive Communications and Computing, University of California, Irvine, CA, USA (e-mail: rangl2@uci.edu).}
\thanks{Q. Liu is with the School of Computer Science and Technology, Dalian University of Technology, Dalian 116024, China (e-mail: qianliu@dlut.edu.cn).}
}
\author{Peng Jiang, Ming Li,~\IEEEmembership{Senior Member,~IEEE}, Rang Liu,~\IEEEmembership{Member,~IEEE}, Wei Wang,~\IEEEmembership{Senior Member,~IEEE}, and Qian Liu,~\IEEEmembership{Member,~IEEE}}
\begin{document}
\maketitle
\thispagestyle{empty}
\begin{abstract}
Integrated Sensing and Communication (ISAC) has emerged as a key enabler for future wireless systems. The recently developed symbol-level precoding (SLP) technique holds significant potential for ISAC waveform design, as it leverages both temporal and spatial degrees of freedom (DoFs) to enhance multi-user communication and radar sensing capabilities. Concurrently, reconfigurable intelligent surfaces (RIS) offer additional controllable propagation paths, further amplifying interest in their application. However, previous studies have encountered substantial computational challenges due to the complexity of jointly designing SLP-based waveforms and RIS passive beamforming. In this paper, we propose a novel model-driven learning approach that jointly optimizes waveform and beamforming by unfolding the iterative alternative direction method of multipliers (ADMM) algorithm.
Two joint design algorithms are developed for radar target detection and direction-of-arrival (DoA) estimation tasks in a cluttered RIS-ISAC system. While ensuring the communication quality-of-service (QoS) requirements, our objectives are: 1) to maximize the radar output signal-to-interference-plus-noise ratio (SINR) for target detection, and 2) to minimize the Cram\'{e}r-Rao bound (CRB) for DoA estimation. Simulation results verify that our proposed model-driven learning algorithms achieve satisfactory communication and sensing performance, while also offering a substantial reduction in computational complexity, as reflected by the average execution time.
\end{abstract}
\vspace{-0.0 cm}
\section{Introduction}

With the rapid advancement of the Internet of Things (IoT), autonomous driving, and other emerging technologies, integrated sensing and communication (ISAC) is becoming a pivotal component of next-generation 6G wireless networks \cite{CuiWC2024}, \cite{AK2024}. In order to reap the benefits of integration and coordination gains in ISAC systems, it is essential to carefully consider the inherent conflicts between communication and sensing requirements in order to achieve satisfactory trade-offs. Towards this goal, the reconfigurable intelligent surface (RIS)-enabled ISAC system has attracted increasing attention in recent years \cite{Survey}, \cite{LRWC2023}. In challenging environments, such as urban areas with dense obstacles or non-line-of-sight scenarios, RIS ensures reliable connectivity and sensing \cite{RIS_add1}. Its lightweight and scalable nature allows for cost-effective deployment, even in areas where traditional infrastructure is impractical. Furthermore, RIS seamlessly integrates communication and sensing functions, dynamically balancing resources to adapt to different application requirements \cite{RIS_ISAC}. By dynamically reconfiguring the electromagnetic properties of its elements, RIS can shape and reflect signals in a way that intelligently manipulates the wireless propagation environment to improve coverage, suppress interference, and boost energy efficiency in ISAC systems \cite{ZQ2022}. This flexibility ensures it meets the diverse needs of modern wireless systems, making RIS an indispensable component of RIS-ISAC systems.

Although RIS-enabled ISAC systems can improve communication and sensing performances across various applications by creating additional controllable propagation paths, it remains crucial to focus on the waveform design at the transmitter side. This is necessary to achieve the spatial and temporal properties required for radar sensing, as well as to obtain beamforming gains and multiplexing for communications \cite{LiuM2023}. By carefully designing the dual-functional waveform, the RIS-ISAC system can further leverage the spatial degrees of freedom (DoFs) provided by RIS, thereby improving both communication and sensing performance \cite{ZWSurvey2022}. Therefore, the transmit dual-functional waveform design problem has recently sparked increasing concern \cite{LF2018}, \cite{BT2023}.

Previous research on waveform design for ISAC systems \cite{LRTWC2024}-\cite{add_re4} has mainly focused on optimizing spatial characteristics through beamforming, using communication metrics such as achievable rate \cite{LRTWC2024}, signal-to-interference-plus-noise ratio (SINR), and multi-user interference (MUI)  \cite{CW2023}, and radar metrics like radar-only precoder similarity \cite{add_ref}, and mean squared error (MSE) \cite{add_re4}. However, these approaches rely on block-level precoding, which exploits only spatial DoFs and cannot fully leverage temporal DoFs due to linear processing and second-order statistics \cite{LASurvey2023}. While block-level precoding/beamforming designs are effective for communication systems transmitting random symbols, radar systems require deterministic detection sequences with favorable temporal characteristics. Consequently, block-level precoding fails to ensure superior sensing performance in ISAC systems.

Recently, symbol-level precoding (SLP)-based waveform design has gained significant attention in ISAC systems \cite{LFGCWorkshop2023}-\cite{LRJSTSP2022}. Unlike traditional block-level precoding, SLP is an advanced nonlinear precoding approach. It allows for the meticulous design of each transmission symbol within each time-slot to meet quality of service (QoS) requirements for communication, while also enhancing radar sensing performance by carefully designing the waveform across symbol time-slots. By exploiting additional temporal DoFs to design transmitter-side waveform, SLP has demonstrated its substantial potential in improving both communication and sensing performance in ISAC systems \cite{LRJSAC}.

Despite the performance gains, the significant design complexity introduced by RIS technologies and SLP-based waveform design hinders their practical applications. On the one hand, compared to traditional block-level precoding methods, SLP introduces additional temporal dimension in precoding design, leading to a substantial computational burden. On the other hand, although RIS enhances ISAC system performance by manipulating the wireless propagation environment, the additional spatial degrees of freedom provided by RIS passive beamforming design also pose challenges to computational complexity \cite{LRJSTSP2022}. Thus, before SLP and RIS can be practically deployed in ISAC systems, it is essential to explore lightweight methods for the joint design of dual-functional waveforms and RIS passive beamforming.

To address these challenges, deep learning techniques offer promising solutions for the joint precoding designs in ISAC systems \cite{add_re1}, \cite{QQ2023}. Recent studies have made initial progress in applying deep learning methods to ISAC systems \cite{JP2024}-\cite{add_re7}, highlighting its importance as a crucial technology for the implementation of 6G ISAC systems. Based on the extensive research, deep learning methods can be broadly categorized into data-driven and model-driven learning schemes \cite{YWMagazine2022}. Data-driven schemes rely on large amounts of historical data for neural model training \cite{Bo2023}, while model-driven approaches combine theoretical frameworks from traditional optimization algorithms with deep learning models \cite{ZhangTWC2023}. However, existing work on data-driven learning-assisted waveform and beamforming design remains limited by the datasets tailored to fixed scenarios \cite{CM1}. To address this challenge, model-driven learning strategies have garnered increasing attention for their potential to enhance the generalization capabilities \cite{CM2}. In specific, the deep unfolding networks (DUNs) allows model-driven learning approaches to transform iterative optimization algorithms into learning tasks, thereby reducing the complexity of operations such as large-scale iterations and matrix inversions \cite{WJ2024}, \cite{JWTC2024}. Consequently, the deployment of model-driven learning methods in RIS-assisted ISAC systems offers a promising approach for enhancing both radar and communication performance, meriting further investigation.

Building on the above discussion, this paper aims to realize low-complexity joint design of SLP-based dual-functional waveforms and RIS passive beamforming in ISAC systems by leveraging innovative model-driven learning techniques. The primary contributions of our research can be summarized as follows:
\begin{itemize}
\item
This paper investigates model-driven learning approaches for achieving low-complexity joint designs of SLP-based dual-functional waveforms and RIS passive beamforming. We consider both radar target detection and direction of arrival (DoA) estimation tasks in a clutter-rich RIS-ISAC system. Radar output SINR and the Cram\'{e}-Rao bound (CRB)  are respectively employed as radar performance metrics to formulate optimization problems. We introduce an innovative model-driven learning framework to tackle these complex non-convex problems due to the complicated radar performance metric and the constraints of constructive interference (CI)-based communication QoS, transmission power budget, and the unit modulus of the RIS reflection coefficients.

\item For the target detection task, a model-driven learning algorithm is developed by leveraging unfolded alternative direction method of multiplier (ADMM), augmented Lagrangian method (ALM), and Davidon-Fletcher-Powell (DFP) optimization techniques to tackle the complicated non-convex SINR maximization problem. Simulation results show that, compared to conventional convex optimization approaches, the proposed ADMM-ALM-DFP model-driven learning algorithm achieves similar radar detection and communication performance while significantly reducing the execution time in the joint waveform and beamforming design.

\item For the DoA estimation task, we also employ the proposed model-driven learning framework to decompose the complex CRB minimization problem into several more tractable sub-problems by utilizing an ADMM-based unfolding technique. In particular, the Powell-Hestenes-Rockafellar (PHR) transformation is introduced to address the highly nonlinear DoA estimation problem with multiple inequality constraints, thereby facilitating robust solutions and enhancing convergence capabilities.
Simulation results demonstrate that, compared to the unsupervised learning scheme, the proposed ADMM-PHR-ALM-DFP model-driven learning algorithm achieves superior performance in both DoA estimation and communication, while maintaining a comparable average execution time. Furthermore, the results demonstrate that with the incorporation of unfolded PHR-ALM, our proposed model-driven framework exhibits enhanced convergence capabilities relative to the unfolded ALM scheme.
\end{itemize}

\textit{Notation}: Scalar variables are denoted by normal-face letters $a$, while vectors and matrices are denoted by lower and upper case letters, $\mathbf{a}$ and $\mathbf{A}$, respectively. $|a|$ and $\|\mathbf{a}\|$ are the absolute value of a scalar $a$ and the norm of a vector $\mathbf{a}$, respectively.
$\mathbb{C}$ and $\mathbb{R}$ denote the set of complex numbers and real numbers, respectively. For a matrix $\mathbf{A}$, ${(\cdot)}^T$ and ${(\cdot)}^H$ denote its transpose and conjugate transpose, respectively. $\mathbf{a}^{(i)}$ denotes the $i$-th iterative solution of $\mathbf{a}$. The angle of complex-valued ${a}$ is presented as $\angle{a}$. The real part and imaginary part of $a$ is expressed by $\Re\{\cdot\}$ and $\Im\{\cdot\}$, respectively. ${\nabla}_{\mathbf{a}}$ is the gradient operator of variable $\mathbf{a}$. $\otimes$ denotes the Kronecker product.

\vspace{-0.0 cm}

\section{System Model}
\label{SystemModel}
\begin{figure}[!t]
\centering
\includegraphics[width=3.51in]{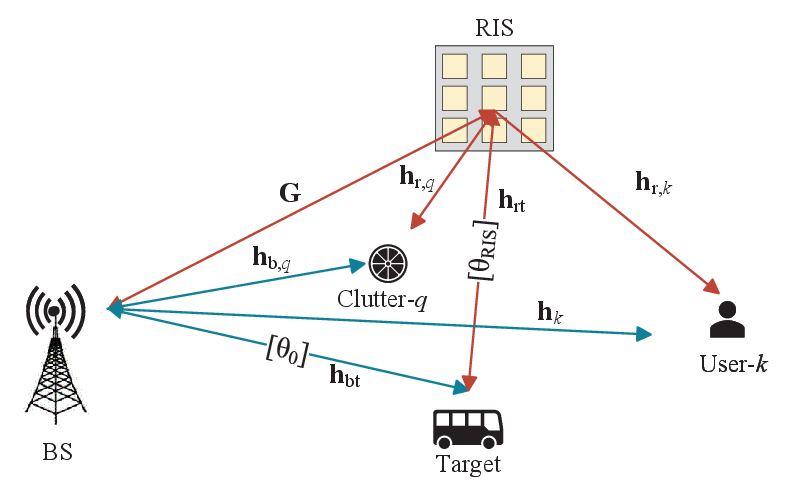}
\caption{RIS-ISAC system.}
\label{system}
\end{figure}

In this paper, we consider an RIS-aided ISAC system as shown in Fig. \ref{system}, where the base station (BS) equipped with $M$ transmit/receive antennas simultaneously serve $K$ communication users and senses a potential target. Particularly, we assume that the ISAC-BS transmits pulse-modulated signals and then receives the echo signals under the time-division (TD) mode. An $N$-element RIS is deployed to support the ISAC system.
The BS jointly designs the SLP-based dual-functional transmit waveform, RIS passive beamforming, and receive filter to suppress $Q$ clutters and achieve better sensing performance while satisfying multi-user communication QoS.

\subsection{Multi-user Communication Model}
For multi-user communication function, we define the symbol information transmitted to the $K$ communication users in the $l$-th time-slot (symbol duration) as $\mathbf{s}[l]\triangleq\left[s_{1}[l],s_2[l],\ldots,s_{K}[l]\right]^{T}$, where $s_{k}[l]$ is independently selected from the $\Omega$-phase shift keying (PSK) constellation map. Through the SLP-based design, each symbol vector $\mathbf{s}[l]$ is precoded to $\mathbf{x}[l]\triangleq[x_{1}[l], x_{2}[l], \ldots, x_{M}[l]]^T\in\mathbb{C}^M$, which is the baseband signal to be transmitted by $M$ antennas.
This precoding is a nonlinear operation applied to each symbol vector, hence it is referred to as symbol-level precoding.
The RIS reflecting coefficients are denoted as $\bm{\phi}\triangleq[{\phi}_1,\phi_2, \ldots,{\phi}_{N}]^T$, with $\left|{\phi}_{n}\right|=1,\forall n\in\left\{1,2,\ldots,N\right\}$. Accordingly, the signal received at the $k$-th user in the $l$-th time-slot can be expressed as
\begin{align}
y_{k}[l]=(\mathbf{h}_{k}^{H}+\mathbf{h}_{\rm r,\it k}^{H}\bm{\Phi}\mathbf{G})\mathbf{x}[l]+n_{k}[l],
\end{align}
where $\bm{\Phi}={\rm diag}\left\{\bm{\phi}\right\}$, $\mathbf{G}\in\mathbb{C}^{N \times M}$, $\mathbf{h}_{k}\in\mathbb{C}^{M}$ and $\mathbf{h}_{\rm r,\it k}\in\mathbb{C}^{N}$ denote the channels between the BS and the RIS, between the BS and the $k$-th user, and between the RIS and the $k$-th user, respectively. The scalar $n_{k}[l]$ represents the additive white Gaussian noise (AWGN) for the $k$-th user and ${n}_{k}[l]\sim\mathcal{CN}(0,{\sigma}_{k}^{2})$.

\begin{figure}[!t]
\centering
\includegraphics[width= 2.6 in]{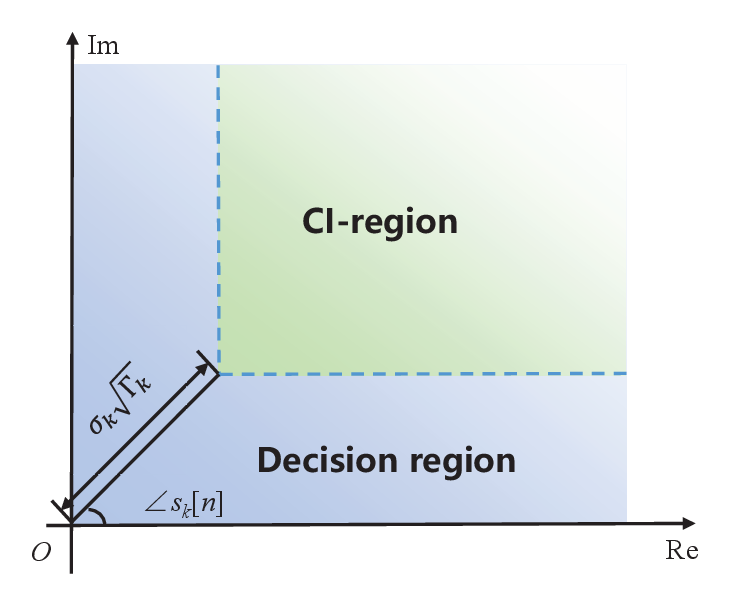}
 \caption{Constructive region and decision region for a specific QPSK symbol.}
 \label{SLP}
 \vspace{0.0cm}
 \end{figure}

The use of SLP in multi-user communication systems can transform the harmful MUI into a beneficial signal by using the concept of CI, therefore strengthening the symbol error rate (SER) performance and  improving communication QoS. Thus, the Euclidean distance between the received noise-free signal and its nearest decision boundary can be used as the multi-user communication QoS metric \cite{LRJSTSP2022} since it directly determines the SER. As shown in Fig. \ref{SLP}, we adopt the quadrature phase shift keying (QPSK) modulated signal as an example to illustrate the CI-based communication QoS metric. For the transmitted symbol $s_{k}[l] = e^{\jmath\pi/4}$ of $k$-th user in $l$-th time-slot, the blue sector and the green sector correspond to the decision region and the CI-region, respectively. Specifically, in order to demodulate the symbols satisfying a required SER performance, the noise-free signal $(\mathbf{h}_{k}^{H}+\mathbf{h}_{\rm r,\it k}^{H}\bm{\Phi}\mathbf{G})\mathbf{x}[l]$ needs to fall into the CI region. Therefore, for the $k$-th communication user in $l$-th time-slot, the communication QoS metric can be concisely expressed as \cite{LRJSAC}
\begin{align}
\label{QoS}
 &\Re\{(\mathbf{h}_{k}^{H}+\mathbf{h}_{\rm{r},\it{k}}^{H}\bm{\Phi}\mathbf{G})\mathbf{x}[l]e^{-\jmath{\angle}s_{k}[l]}\}\sin\Theta \\
 &~~-|\Im\{(\mathbf{h}_{k}^{H}+\mathbf{h}_{\rm r,\it k}^{H}\bm{\Phi}\mathbf{G})\mathbf{x}[l]e^{-\jmath{\angle}s_{k}[l]}\}|\cos\Theta\geq{\gamma}_{k,l},\forall k,l,\nonumber
\end{align}
where $\Theta=\pi/4$ under the QPSK modulation, ${\gamma}_{k,l}\triangleq{\sigma}_{k}\sqrt{{\Gamma}_{k}}\sin\Theta$ represents the preset minimum communication requirement, and ${\Gamma}_{k}$ denotes the required signal-to-noise ratio (SNR) for the $k$-th communication user.

\subsection{Radar Sensing Model}
While carrying symbol information for $K$ users, the dual-functional waveform simultaneously fulfills a sensing role, enabling the detection and estimation of potential targets. Generally, the target echo signal at the radar receiver of the BS consists of the signals from four main paths, BS-target-BS, BS-target-RIS-BS, BS-RIS-target-BS, and BS-RIS-target-RIS-BS.
For brevity, we represent the echo signal corresponding to the cell under test (i.e., the range cell of the target) for the transmitted signal $\mathbf{x}[l]$ as $\mathbf{r}_{\rm t}[l]$. Thus, the target echo signal is expressed as
\be
 \mathbf{r}_{\rm t}[l]={\alpha}_{0}(\mathbf{h}_{\rm bt}+\mathbf{G}^H\bm{\Phi}\mathbf{h}_{\rm rt})(\mathbf{h}_{\rm bt}^H+\mathbf{h}_{\rm rt}^H\bm{\Phi}\mathbf{G})\mathbf{x}[l]e^{\jmath2\pi(l-1)\upsilon_0},
\ee
where ${\alpha}_0$ is the radar cross section (RCS) of the target with $\mathbb{E}\{|{\alpha}_0|^2\}={\xi}_0^2$, $\upsilon_{0}$ denotes the target Doppler frequency. $\mathbf{h}_{\rm bt}\in\mathbb{C}^M$ and $\mathbf{h}_{\rm rt}\in\mathbb{C}^N$ represent the line-of-sight (LoS) channels between the BS and the target and between the RIS and the target, which are determined by the DoA with respect to the BS $\theta_{0}$ and the RIS $\theta_{\rm RIS}$, respectively.
The interference signal caused by clutter follows propagation paths similar to those of the target echo signal. In this paper, we assume that the BS can obtain the knowledge of the range-angle positions and the mean power of the clutter sources from a cognitive paradigm. Generally, by analyzing an environmental dynamic database including previous scanning/tracking files, geographical information system, and clutter models, the BS can get the side-information in practice to further guide the waveform and passive beamforming design. Accordingly, the range-angle location of the $q$-th clutter source denoted as $(d_q,{\theta}_q)$, $d_q \in \{0,\ldots,L\}$.
Thus, the interference signal $\mathbf{r}_{\rm c}[l]$ can be expressed as
\begin{align}
 \mathbf{r}_{\rm c}[l]=&\sum_{q=1}^{Q}{\alpha}_{q}(\mathbf{h}_{\rm b,\it q}\!+\!\mathbf{G}^{H}\bm{\Phi}\mathbf{h}_{\rm r,\it q})\nonumber\\&\times(\mathbf{h}_{\rm b,\it q}^{H}\!+\!\mathbf{h}_{\rm r,\it q}^{H}\bm{\Phi}\mathbf{G})\mathbf{x}[l\!-\!d_q]e^{\jmath2\pi(l-1)\upsilon_q},
\end{align}
where $\alpha_q$ denotes the RCS of the $q$-th clutter with $\mathbb{E}\{|{\alpha}_q|^2\}={\xi}_q^2$, $\upsilon_{q}$ represents the Doppler frequency of the $q$-th clutter. The notations $\mathbf{h}_{\rm b,\it q}\in\mathbb{C}^{M}$ and $\mathbf{h}_{\rm r,\it q}\in\mathbb{C}^{N}$ denote the LoS channels between the BS and the $q$-th clutter and between the RIS and the $q$-th clutter. Since both the target and the clutter are slowly moving in the considered RIS-ISAC scenarios, the Doppler frequencies are assumed that $\upsilon_0 = \upsilon_q = 0, \forall q$. Thus, the baseband echo signal received by the BS in the $l$-th time-slot can be defined as
\begin{align}
\label{signal_model}
 \mathbf{r}[l]=\mathbf{r}_{\rm t}[l]+\mathbf{r}_{\rm c}[l]+\mathbf{z}[l],
\end{align}
where $\mathbf{z}[l]\sim\mathcal{CN}(0,{\xi}_{\rm z}^{2}\mathbf{I}_{M})$ denotes the AWGN. For conciseness, the complex cascade channel is denoted as
\begin{subequations}
\begin{align}
\mathbf{H}_{0}(\bm{\phi})&\triangleq(\mathbf{h}_{\rm bt}+\mathbf{G}^{H}\bm{\Phi}\mathbf{h}_{\rm rt})(\mathbf{h}_{\rm bt}^{H}+\mathbf{h}_{\rm rt}^{H}\bm{\Phi}\mathbf{G}),\\
\mathbf{H}_{q}(\bm{\phi})&\triangleq(\mathbf{h}_{\rm b,\it q}+\mathbf{G}^{H}\bm{\Phi}\mathbf{h}_{\rm r,\it q})(\mathbf{h}_{\rm b,\it q}^{H}+\mathbf{h}_{\rm r,\it q}^{H}\bm{\Phi}\mathbf{G}),
\end{align}
\end{subequations}
then the received signal can be rewritten as
\be
\label{signal_model_1}
 \mathbf{r}[l]= {\alpha}_{0}\mathbf{H}_{0}(\bm{\phi})\mathbf{x}[l]+\sum_{q=1}^{Q}{\alpha}_{q}\mathbf{H}_{q}(\bm{\phi})\mathbf{x}[l-d_q]+\mathbf{z}[l].
\ee

For the radar sensing function in this ISAC system, each radar pulse has $L$ digital samples (i.e. time-slots) and the range domain is divided into $L$ discrete bins. In processing the echo signal of each radar pulse, the receiver stacks the received signal as
$\mathbf{r}\triangleq[\mathbf{r}[1]^T,\mathbf{r}[2]^T,\ldots,\mathbf{r}[L]^T]^T$. For facilitating the algorithm development, we also represent the dual-functional waveform for $L$ time-slots as a vector form: $\mathbf{x}\triangleq[\mathbf{x}[1]^T,\mathbf{x}[2]^T,\ldots,\mathbf{x}[L]^T]^T\in\mathbb{C}^{ML}$.
Accordingly, the received signals of $L$ samples can be formulated as
\begin{align}
\label{stack_signal}
 \mathbf{r}=\alpha_{0}\mathbf{\widetilde{H}}_{0}(\bm{\phi})\mathbf{x}+\sum_{q=1}^{Q}{\alpha}_{q}\mathbf{\widetilde{H}}_{q}(\bm{\phi})\mathbf{x}+\mathbf{z},
\end{align}
where  $\mathbf{z}\triangleq[\mathbf{z}[1]^T,\mathbf{z}[2]^T,\ldots,\mathbf{z}[L]^T]^T$, $\mathbf{\widetilde{H}}_{0}(\bm{\phi}) \triangleq \mathbf{I}_{L}\otimes\mathbf{H}_{0}(\bm{\phi})$ and $\mathbf{\widetilde{H}}_{q} \triangleq {[\mathbf{I}_{L}\otimes\mathbf{H}_{q}(\bm{\phi})]}\mathbf{J}_{q}$ for brief expression. Specially, in order to represent the propagation delay of clutter returns from different ranges, the shift matrix for $q$-th clutter $\mathbf{J}_{q}$ is introduced, and the $(i,j)$-th element of $\mathbf{J}_{q}$ is defined as
\begin{align}
\mathbf{J}_{q}(i,j)=
\begin{cases}
 1, ~i-j=Md_q,\\
0,~\textrm{otherwise}.
\end{cases}
\end{align}

After receiving the echo signal (\ref{stack_signal}), the BS can realize the sensing function by performing appropriate  radar signal processing.  Typical radar sensing tasks include target detection and parameter estimation.
In order to develop a comprehensive RIS-ISAC system capable of multi-task, the following two sections present the joint dual-functional waveform and RIS passive beamforming designs for target detection and DoA estimation tasks, respectively.

\section{Joint Waveform and Passive Beamforming Design for Target Detection}
\label{Sec:detection}
In this section, we will focus on target detection task in the RIS-ISAC system and introduce a joint dual-functional waveform and passive beamforming design to achieve superior target detection performance while maintaining satisfactory multi-user communication QoS.
To achieve better taget detection performance, the stacked radar echo signals are processed through a linear space-time receive filter $\mathbf{w}\in\mathbb{C}^{ML}$, whose output can be written as
\begin{align}
   r_{\rm o}=  \mathbf{w}^{H}\mathbf{r}.
\end{align}
Thus, the hypothesis testing problem for the radar output can be expressed as
\begin{equation}
r_{\rm o}=
\begin{cases}
 \mathcal{H}_{0}:\sum_{q=1}^{Q}{\alpha}_{q}\mathbf{w}^{H}\widetilde{\mathbf{H}}_{q}(\bm{\phi})\mathbf{x}+\mathbf{w}^{H}\mathbf{z},\\
\mathcal{H}_{1}: {\alpha}_{0}\mathbf{w}^{H}\widetilde{\mathbf{H}}_{0}(\bm{\phi})\mathbf{x}+\sum_{q=1}^{Q}{\alpha}_{q}\mathbf{w}^{H}\widetilde{\mathbf{H}}_{q}(\bm{\phi})\mathbf{x}+\mathbf{w}^{H}\mathbf{z}.
\end{cases}
\end{equation}
The conditional probability distributions of $r_{\rm o}$ can be expressed as $r_{\rm o}|\mathcal{H}_{0} \sim \mathcal{CN}(0, {\epsilon}_{0})$ and $r_{\rm o}|\mathcal{H}_{1}\sim\mathcal{CN}(0, {\epsilon}_{1})$, where we define
\begin{subequations}\begin{align}
    \epsilon_{0}&\triangleq{\xi}_{\rm z}^{2}\mathbf{w}^{H}\mathbf{w}+\sum_{q=1}^{Q}{\xi}_{q}^{2}\mathbf{w}^{H}\widetilde{\mathbf{H}}_{q}(\bm{\phi})\mathbf{x}\mathbf{x}^{H}\widetilde{\mathbf{H}}_{q}^{H}(\bm{\phi})\mathbf{w},\\
    \epsilon_{1}&\triangleq\epsilon_{0}+{\xi}_{0}^{2}\mathbf{w}^{H}\widetilde{\mathbf{H}}_{0}(\bm{\phi})\mathbf{x}\mathbf{x}^{H}\widetilde{\mathbf{H}}_{0}^{H}(\bm{\phi})\mathbf{w}.
\end{align}\end{subequations}
Then, the Neyman-Pearson detector \cite{DT1} is deployed to identify whether the target is present by: \be \mathcal{T} = {|r_{\rm o}|}^{2} \mathop{\gtrless}\limits_{\mathcal{H}_{0}}^{\mathcal{H}_{1}} \delta ,\ee  where $\delta$ is the decision threshold determined by the probability of false alarm. The statistic distribution of $\mathcal{T}$ is thus given as
\begin{align}
     \mathcal{T}\sim
     \begin{cases}
    \epsilon_{0}\chi_{2}^{2}, ~~~\mathcal{H}_{0},\\
    \epsilon_{1}\chi_{2}^{2},~~~\mathcal{H}_{1},
\end{cases}
\end{align}
where $\chi_{2}^{2}$ is central chi-square distribution with two DoFs.
Given the decision threshold, the probabilities of detection $P_{\rm d}$ and false alarm $P_{\rm fa}$ can be respectively given as \cite{DT2}
\begin{subequations}\begin{align}
P_{\rm d}={\rm Pr}(\mathcal{T}>\delta|\mathcal{H}_{1})=1-\mathcal{F}_{\chi_{2}^{2}}(\delta/\epsilon_{1}),\label{eq: pd}\\
P_{\rm fa}={\rm Pr}(\mathcal{T}>\delta|\mathcal{H}_{0})=1-\mathcal{F}_{\chi_{2}^{2}}(\delta/\epsilon_{0}),\label{eq: pfa}
\end{align}\end{subequations}
where $\mathcal{F}_{\chi_{2}^{2}}(\cdot)$ represents the central chi-squared distribution function with two DoFs and ${\rm Pr}(\cdot)$ is the probability function. Substituting (\ref{eq: pfa}) into (\ref{eq: pd}), the probability of detection $P_{\rm d}$ can be obtained as
\begin{align}
    P_{\rm d}=1-\mathcal{F}_{\chi_{2}^{2}}\Big(\frac{\epsilon_{0}}{\epsilon_{1}}\mathcal{F}_{{\chi}_{2}^{2}}^{-1}(1-P_{\rm fa})\Big)\propto \frac{\epsilon_{0}}{\epsilon_{1}}=\textrm{SINR}+1,
\end{align}
where the radar output SINR can be expressed as
\begin{align}
\textrm{SINR}=\frac{{\xi}_{0}^{2}{|\mathbf{w}^{H}\mathbf{\widetilde{H}}_{0}(\bm{\phi})\mathbf{x}|}^2}{\mathbf{w}^{H}\Big[\sum_{q=1}^{Q}{\xi}_{q}^{2}\mathbf{\widetilde{H}}_{q}(\bm{\phi})\mathbf{x}\mathbf{x}^{H}\mathbf{\widetilde{H}}_{q}^{H}(\bm{\phi})
+{\xi}_{z}^{2}\mathbf{I}_{ML}\Big]\mathbf{w}}.
\end{align}

It is noteworthy that the target detection probability is positively proportional to the radar output SINR. Therefore, in this paper, we adopt the radar output SINR as a sensing metric to evaluate the target detection performance. Our goal is to jointly design the dual-functional waveform $\mathbf{x}$, the passive beamforming $\bm{\phi}$, and the receive filter $\mathbf{w}$ to maximize the radar output SINR while satisfying the requirement of multi-user communication QoS and the constant-modulus constraints of waveform and RIS. Thus, the optimization problem is formulated as
\begin{subequations}
\label{object}
\begin{eqnarray}
\max_{\mathbf{w},\mathbf{x},\bm{\phi}}&\ddfrac{{\xi}_{0}^{2}{|\mathbf{w}^{H}\mathbf{\widetilde{H}}_{0}(\bm{\phi})\mathbf{x}|}^2}
{\mathbf{w}^{H}\big[\sum_{q=1}^{Q}{\xi}_{q}^{2}\mathbf{\widetilde{H}}_{q}(\bm{\phi})\mathbf{x}\mathbf{x}^{H}\mathbf{\widetilde{H}}_{q}^{H}(\bm{\phi})
+{\xi}_{z}^{2}\mathbf{I}\big]\mathbf{w}}\\
\rm{s.t.} &g_{k,l}(\mathbf{x}[l],\bm{\phi})\leq0,~\forall k,l, ~~~~~~~~~~~~~~~~~~~~\\
&~ \left|x_{i}\right| = \sqrt{P / M}, ~\forall i~~~~~~~~~~~~~~~~~~~~~~~~~~~\\
& {\left|{\phi}_{n}\right|=1},~~~~~~~~~~~~~~~~~~~~~~~~~~~~~~~~~~~~~
\end{eqnarray}
\end{subequations}
where $P$ represents the power budget at the BS transmitter and the multi-user communication metric $g_{k,l}(\mathbf{x}[l],\bm{\phi})$ is defined according to (\ref{QoS}) as
\begin{align}
g_{k,l}(\mathbf{x},\bm{\phi})&={\gamma}_{k,l}-\Re\{(\mathbf{h}_{k}^{H}+\mathbf{h}_{\rm r, \it k}^{H}\bm{\Phi}\mathbf{G})\mathbf{x}[l]e^{-\jmath{\angle}s_{k}[l]}\}\sin\Theta \nonumber\\
&\quad+|\Im\{(\mathbf{h}_{k}^{H}+\mathbf{h}_{\rm r, \it k}^{H}\bm{\Phi}\mathbf{G})\mathbf{x}[l]e^{-\jmath{\angle}s_{k}[l]}\}|\cos\Theta.
\end{align}

We see that the sub-problem with respect to the receive filter $\mathbf{w}$ can be transformed to a minimum variance distortionless response (MVDR) problem given $\mathbf{x}$ and $\bm{\phi}$. Accordingly, the optimal receive filter ${\mathbf{w}}^{\star}$ can be obtained by

\begin{small}\begin{align}
{\mathbf{w}}^{\star}=\frac{\big[\sum_{q=1}^{Q}{\xi}_{q}^{2}\mathbf{\widetilde{H}}_{q}(\bm{\phi})\mathbf{x}\mathbf{x}^{H}\mathbf{\widetilde{H}}_{q}^{H}(\bm{\phi})
\!+\!{\xi}_{z}^{2}\mathbf{I}\big]^{-1}\mathbf{\widetilde{H}}_{0}(\bm{\phi})\mathbf{x}}
{\mathbf{x}^{H}\mathbf{\widetilde{H}}_{0}^{H}(\bm{\phi}){\big[\sum_{q=1}^{Q}{\xi}_{q}^{2}\mathbf{\widetilde{H}}_{q}(\bm{\phi})\mathbf{x}\mathbf{x}^{H}\mathbf{\widetilde{H}}_{q}^{H}(\bm{\phi})
+{\xi}_{z}^{2}\mathbf{I}\big]}^{-1}\mathbf{\widetilde{H}}_{0}(\bm{\phi})\mathbf{x}}.
\end{align}\end{small}

\noindent Substituting $\mathbf{w}^\star$ into the objective function (\ref{object}a), the original problem can be reformulated as
\begin{subequations}
\label{problem}
\begin{align}
&\min_{\mathbf{x},\bm{\phi}}~\mathbf{x}^{H}\mathbf{\widetilde{H}}_{0}^{H}(\bm{\phi})\Big[\sum_{q=1}^{Q}\!{\xi}_{q}^{2}\mathbf{\widetilde{H}}_{q}(\mathbf{\bm{\phi}})\mathbf{x}\mathbf{x}^{H}\mathbf{\widetilde{H}}_{q}^{H}(\bm{\phi})
\!+\!{\xi}_{z}^{2}\mathbf{I}\Big]^{-1}\mathbf{\widetilde{H}}_{0}(\bm{\phi})\mathbf{x}\label{objective}\\
&~~\text{s.t.}~~ g_{k,l}(\mathbf{x},\bm{\phi})\leq0,~\forall k,l, \label{C:QoS}\\
&\qquad~\left|x_{i}\right| = \sqrt{P / M}, ~\forall i,\label{P:x}\\
&\qquad~ {\left|{\phi}_{n}\right|=1}\label{P:phi}.
\end{align}
\end{subequations}

To handle the non-convex and bivariate objective function (\ref{objective}), previous research usually relies on complex transformations, e.g., the majorization-minimization (MM) method \cite{MM}. However, these complex transformations lead to high-dimensional matrix computations and require numerous iterations, resulting in significantly increased computational complexity.
In the remainder of this section, we propose a novel model-driven learning framework based on an unfolded ADMM-ALM-DFP approach to efficiently solve this complicated non-convex problem and reduce the practical execution time.

\subsection{Model-driven Learning Framework}
\begin{figure*}[!t]
\centering
\includegraphics[width=0.9\linewidth]{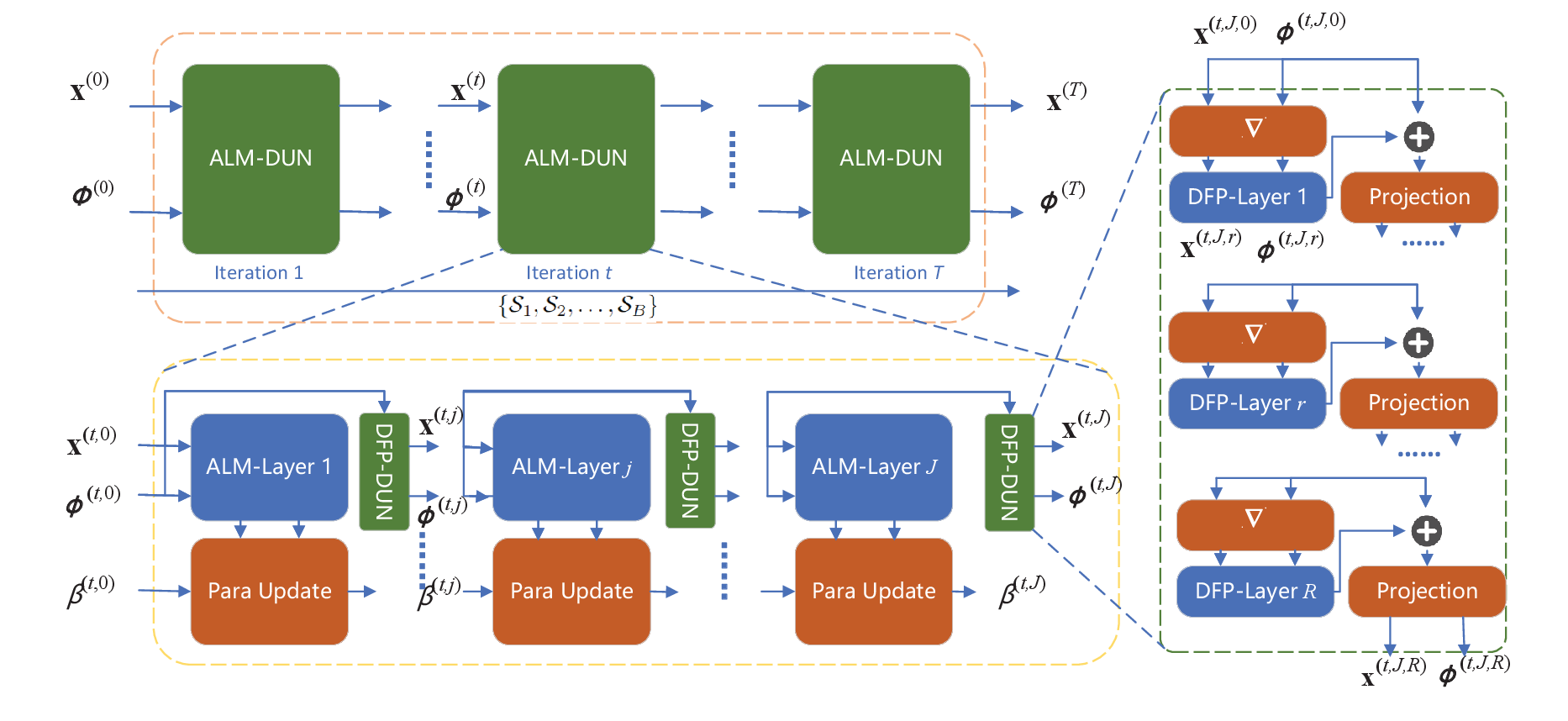}
\caption{Model-driven learning framework.}
\label{Fig:framework}
\vspace{-0.2 cm}
\end{figure*}

The main idea of our proposed model-driven learning framework is to decompose the complex problem into a series of iterative sub-problems that can be solved by unfolding-friendly algorithms, while the hyperparameters of these unfolding-friendly algorithms are learned through DUN, as shown in Fig. \ref{Fig:framework}.
In particular, to deal with the bivariate objective and transform the optimization problem into an unfolding-friendly form, an ADMM-based unfolding approach is introduced in our proposed model-driven learning framework. Meanwhile, we address the communication QoS constraints by training the dual variables of unfolded ALM, and tackle the challenges of the non-convex objective function (\ref{objective}) by training the step size of unfolded-DFP algorithm.

Specifically, by introducing auxiliary variables $\mathbf{v}\triangleq[v_1,v_2,\ldots,v_{ML}]^T$ and $\bm{\varphi}\triangleq[{\varphi}_{1},\varphi_2,\ldots,{\varphi}_{N}]^T$, the joint waveform and beamforming design for target detection task can be transformed to
\begin{subequations}
\label{ADMM}
\begin{align}
\min_{\mathbf{x},\bm{\phi},\mathbf{v},\bm{\varphi}}~&f_{1}(\mathbf{x},\bm{\phi}) \label{SINR_obj}\\
\rm{s.t.} ~~~& g_{k,l}(\mathbf{x},\bm{\phi})\leq0,~\forall k,l, \label{C:QoS2}  ~~~~~~ \\
& \left|x_{i}\right|\leq\sqrt{P/M},~ \forall i, \label{C1}\\
&~{\left|{\phi}_{n}\right|\leq1},~ \forall n, \label{C2}\\
&~\mathbf{v}=\mathbf{x}, \\
&~\bm{\varphi}=\bm{\phi}, \\
& \left|v_{i}\right|=\sqrt{P/M}, ~\forall i, \label{C:v}\\
& ~{\left|{\varphi}_{n}\right|=1},~\forall n, \label{C:varphi}
\end{align}
\end{subequations}
where for brevity we define the objective function (\ref{SINR_obj}) as
\begin{align}
\label{SINR_define}
&f_{1}(\mathbf{x},\bm{\phi})=\nonumber\\
&\mathbf{x}^{H}\mathbf{\widetilde{H}}_{0}^{H}(\bm{\phi})\big[\sum_{q=1}^{Q}{\xi}_{q}^{2}\mathbf{\widetilde{H}}_{q}(\bm{\phi})\mathbf{x}\mathbf{x}^{H}\mathbf{\widetilde{H}}_{q}^{H}(\bm{\phi})
+{\xi}_{z}^{2}\mathbf{I}\big]^{-1}\mathbf{\widetilde{H}}_{0}(\bm{\phi})\mathbf{x}.
\end{align}


To accommodate the ADMM framework for problem (\ref{ADMM}), we introduce an indicator function $\mathbb{I}(\mathbf{x},\bm{\phi},\mathbf{v}, \bm{\varphi})$ to enforce constraints (\ref{C1}), (\ref{C2}), (\ref{C:v}), and (\ref{C:varphi}). In particular, the indicator function applies an infinite penalty when the variables $\mathbf{x}, \bm{\phi}, \mathbf{v}$, and $\bm{\varphi}$ lie outside their feasible regions. Therefore, problem (\ref{ADMM}) is transformed as
\begin{subequations}
\label{Pro:ADMM}
\begin{align}&\min_{\mathbf{x},\bm{\phi},\mathbf{v},\bm{\varphi}}f_{1}(\mathbf{x},\bm{\phi})+\mathbb{I}(\mathbf{x},\bm{\phi},\mathbf{v},\bm{\varphi})+\frac{\rho}{2}\|\mathbf{x}-\mathbf{v}+\frac{\bm{\mu}_{1}}{\rho}\|^2
\\&~~~~~~~~~~~~~~~~+\frac{\rho}{2}\|\bm{\phi}-\bm{\varphi}+\frac{\bm{\mu}_{2}}{\rho}\|^2,\nonumber\\
&\textrm{s. t.}~~~~~g_{k,l}(\mathbf{x}, \bm{\phi})\leq0, \forall k, l,
\end{align}
\end{subequations}
where $\rho>0$ denotes the penalty parameter, and $\bm{\mu}_{1}$ and $\bm{\mu}_{2}$ represent the dual variables introduced by ADMM framework. Then, the ADMM-based problem (\ref{Pro:ADMM}) can be iteratively solved by alternately updating $\mathbf{x}$, $\mathbf{v}$, $\bm{\phi}$, $\bm{\varphi}$, $\bm{\mu}_{1}$, and $\bm{\mu}_{2}$ as shown in the following subsections.

\subsection{Update $\mathbf{x}$}
\label{ADMM_x}
In the $t$-th ADMM iteration, with given $\mathbf{v}^{(t)}$, $\bm{\phi}^{(t)}$, $\bm{\varphi}^{(t)}$, ${\bm{\mu}}_{1}$ and ${\bm{\mu}}_{2}$, we update $\mathbf{x}^{(t)}$ and $\bm{\beta}^{(t)}$ by minimizing the augmented Lagrangian problem in (\ref{Pro:ADMM}). It is noted that the indicator function $\mathbb{I}(\mathbf{x},\bm{\phi}^{(t)},\mathbf{v}^{(t)},\bm{\varphi}^{(t)})$ imposes a bound constraint on $\mathbf{x}$, which can be easily satisfied through normalization. Thus, we temporally ignore the indicator function in the following derivations and formulate the subproblem based on ALM
\begin{align}
\label{eq:x_initial1}
\underset{\mathbf{x},\bm{\beta}}\min~~\mathcal{L}_{1}(\mathbf{x}, \bm{\beta})&=f_{1}(\mathbf{x},\bm{\phi}^{(t)})
+\frac{\rho}{2}\|\mathbf{x}-\mathbf{v}^{(t)}+{\bm{\mu}_{1}}/{\rho}\|^2\nonumber\\
&\quad+\sum_{l=1}^{L}\sum_{k=1}^{K}{\beta}_{k,l}g_{k,l}(\mathbf{x},\bm{\phi}^{(t)}),
\end{align}
where ${\beta}_{k,l}$, $k=1,\ldots, K$, $l=1,\ldots, L$, denote the Lagrangian multiplier. For clearer notation, we also define the vector $\bm{\beta}$ that collects all Lagrangian multipliers as $\bm{\beta}\triangleq[\beta_{1,1},  \ldots, \beta_{K,L}]^T\in\mathbb{R}^{KL}$.

In order to obtain $\mathbf{x}^{(t)}$ in the $t$-th ADMM iteration, the unfolded ALM-DFP method is employed to iteratively update the variables as follows.
In specific, the unfolded ALM method is first applied to iteratively update $\mathbf{x}^{(t)}$ and the Lagrange multiplier $\bm{\beta}^{(t)}$. In the $j$-th unfolded ALM iteration, $\mathbf{x}^{(t,j)}$ and ${\bm{\beta}}^{(t,j)}$ are updated by
\begin{subequations}
\label{x:ALM}
\begin{align}
\beta^{(t,j)}_{k,l}&=\max\{\beta^{(t,j-1)}_{k,l}+{\eta}^{(t,j)}_{k,l}g_{k,l}(\mathbf{x}^{(t,j-1)},\bm{\phi}^{(t)}), 0\} \label{update_beta_x},\\
\mathbf{x}^{(t,j)}&=\arg\min_{\mathbf{x}}~\mathcal{L}_{1}(\mathbf{x}, \bm{\beta}^{(t,j)}),\label{sub1}
\end{align}
\end{subequations}
where ${\eta}_{k,l}^{(t,j)}$, $k=1,\ldots, K$, $l=1,\ldots, L$, denote the learnable step-size parameter for the $j$-th ALM dual update in the $t$-th ADMM iteration.

We notice that the non-convex characterization in (\ref{sub1}) still poses a challenge to obtain $\mathbf{x}^{(t,j)}$. To tackle this issue, we further utilize the unfolding-friendly DFP algorithm to iteratively update $\mathbf{x}^{(t,j)}$ in each ALM loop. Particularly, we employ the generalized Quasi-Newtonian method \cite{BFGS} to solve sub-problem (\ref{sub1}), and thus update $\mathbf{x}$ in the $r$-th iteration by
\begin{align}
\mathbf{x}^{(t,j,r)}=\mathbf{x}^{(t,j,r-1)}-{\tau}^{(t,j,r)}\mathbf{B}^{(r-1)}\mathcal{G}(\mathbf{x}^{(t,j,r-1)}) \label{update_x},
\end{align}
where ${\tau}^{(t,j,r)}$ is the learnable step-size parameter of the DFP algorithm, $\mathcal{G}(\mathbf{x})=\nabla\mathcal{L}_{1}(\mathbf{x})$, and $\mathbf{B}^{(r-1)}$ represents an approximation to the Hessian matrix obtained in the $(r-1)$-th iteration. Then, in the $r$-th unfolded DFP iteration, $\mathbf{B}^{(r)}$ is updated by
\begin{align}
\label{SPDM}
\mathbf{B}^{(r)}=\begin{cases}
\mathbf{B}^{(r-1)},~~\mathrm{if} ~~\frac{(\mathbf{y}^{(r)})^T\bm{\delta}^{(r)}}{\|\bm{\delta}^{(r)}\|_2^2}\leq \|\mathcal{G}(\mathbf{x}^{(t,j,r-1)})\|,\\
\mathbf{B}^{(r-1)}-\widetilde{\mathbf{B}}^{(r-1)} ,~~\textrm{otherwise},
\end{cases}\!
\end{align}
where we define
\begin{subequations}\begin{align}    \widetilde{\mathbf{B}}^{(r-1)} &\triangleq \frac{\mathbf{B}^{(r-1)}\mathbf{y}^{(r-1)}(\mathbf{y}^{(r-1)})^T\mathbf{B}^{(r-1)}}{(\mathbf{y}^{(r-1)})^T\mathbf{B}^{(r-1)}\mathbf{y}^{(r-1)}}\\
    &-\frac{{({\mathbf{y}^{(r-1)}})^T\mathbf{B}^{(r-1)}\mathbf{y}^{(r-1)}}}{((\mathbf{y}^{(r-1)})^T{\bm{\delta}}^{(r-1)})^2}\bm{\delta}^{(r-1)}(\bm{\delta}^{(r-1)})^T,\nonumber\\
    \mathbf{y}^{(r-1)}&\triangleq{\mathcal{G}}(\mathbf{x}^{(t,j,r)})-{\mathcal{G}}(\mathbf{x}^{(t,j,r-1)}),\\
    \bm{\delta}^{(r-1)}&\triangleq\mathbf{x}^{(t,j,r)}-\mathbf{x}^{(t,j,r-1)}.
\end{align}\end{subequations}
To ensure that $\mathbf{x}$ satisfies the power constraint (\ref{C1}), the vector $\mathbf{x}^{(t,j,r)}$ is then scaled as
\begin{align}
\label{x:DFP}
\mathbf{{x}}^{(t,j,r)}(i) :=\frac{\sqrt{P/M}\mathbf{x}^{(t,j,r)}(i)}{|\mathbf{x}^{(t,j,r)}(i)|},~\forall i=1,2,\ldots,ML.
\end{align}
\subsection{Update $\mathbf{v}$}
With obtained $\mathbf{x}^{(t)}$, $\bm{\phi}^{(t)}$, $\bm{\varphi}^{(t)}$, ${\bm{\mu}}_{1}$ and ${\bm{\mu}}_{2}$, the closed-form optimal $\mathbf{v}^{(t)}$ is easily calculated as
\begin{align}
\mathbf{v}^{(t)}=\sqrt{P/M}e^{\jmath\angle(\rho\mathbf{x}^{(t)}+\bm{{\mu}}_{1})}\label{update_v}.
\end{align}
\subsection{Update $\bm{\phi}$}
With given $\mathbf{x}^{(t)}$, $\mathbf{v}^{(t)}$, $\bm{\varphi}^{(t)}$, ${\bm{\mu}}_{1}$ and ${\bm{\mu}}_{2}$, we transform the sub-problem in ALM form to update $\bm{\phi}$ by
\begin{align}
\underset{\bm{\phi},\bm{\beta}}\min~~\mathcal{L}_{1}(\bm{\phi}, \bm{\beta})&=f_{1}(\mathbf{x}^{(t)},\bm{\phi})
+\frac{\rho}{2}\|\bm{\phi}-\bm{\varphi}^{(t)}+\frac{\bm{\mu}_{2}}{\rho}\|^2\nonumber\\
&\quad+\sum_{l=1}^{L}\sum_{k=1}^{K}{\beta}_{k,l}g_{k,l}(\mathbf{x}^{(t)},\bm{\phi}),
\end{align}
in which we also temporally ignore the indicator function since it is easy to meet the bound constraint (\ref{C2}) by normalization.

Following the similar process in (\ref{x:ALM})-(\ref{x:DFP}), in the $j$-th ALM iteration, $\bm{\phi}^{(t,j)}$ and ${\bm{\beta}}^{(t,j)}$ are given by
\begin{subequations}\begin{align}
\beta^{(t,j)}_{k,l}&=\max\big\{\beta^{(t,j-1)}_{k,l}+{\kappa}^{(t,j)}_{k,l}g_{k,l}(\mathbf{x}^{(t)},\bm{\phi}^{(t,j-1)}), 0\big\} \label{update_beta_phi},\!\\
\bm{\phi}^{(t,j)}&=\arg\min_{\bm{\phi}}~\mathcal{L}_1(\bm{\phi}, \bm{\beta}^{(t,j)}),\label{sub2}
\end{align}\end{subequations}
where $\bm{\kappa}^{(t,j)}_{k,l}$, $k=1,\ldots, K$, $l=1,\ldots, L$, represent the learnable step-size parameter for the ALM dual update. Then, we employ an unfolding-friendly modified DFP algorithm to solve the non-convex unconstrained optimization problem (\ref{sub2}). Specifically, in the $r$-th iteration of the generalized Quasi-Newtonian method, the variable $\bm{\phi}^{(t,j,r)}$ is updated by
\begin{align}
\bm{\phi}^{(t,j,r)}=\bm{\phi}^{(t,j,r-1)}-{{\zeta}}^{(t,j,r)}\mathbf{D}^{(r-1)}\nabla\mathcal{L}_{1}(\bm{\phi}^{(t,j,r-1)})\label{update_phi},
\end{align}
where ${\zeta}^{(t,j,r)}$ denotes the learnable step-size parameter for DFP algorithm, $\mathbf{D}^{(r-1)}$ denotes the approximate Hessian matrix obtained in the $(r-1)$-th iteration, which can be updated similarly to $\mathbf{B}^{(r)}$ mentioned in (\ref{SPDM}).
To ensure that the solution $\bm{\phi}$ satisfies the constraint (\ref{C2}), $\bm{\phi}^{(t,j,r)}$ is then scaled as
\begin{align}
\label{phi:DFP}
\bm{\phi}^{(t,j,r)}(n):=\frac{{\bm{\phi}^{(t,j,r)}}(n)}{|\bm{\phi}^{(t,j,r)}(n)|},~\forall n = 1,2,\ldots,N.
\end{align}
\subsection{Update $\bm{\varphi}$}
With $\mathbf{x}^{(t)}$, $\mathbf{v}^{(t)}$, $\bm{\phi}^{(t)}$, ${\bm{\mu}}_{1}$ and ${\bm{\mu}}_{2}$, the closed-form optimal $\bm{\varphi}^{(t)}$ is easily expressed as
\begin{align}
\bm{\varphi}^{(t)} =e^{\jmath\angle(\rho\bm{\phi}^{(t)}+\bm{{\mu}}_{2})}\label{update_varphi}.
\end{align}
\subsection{Update $\bm{\mu}$}
\label{ADMM_mu}
After obtaining $\mathbf{x}^{(t)}$, $\mathbf{v}^{(t)}$, $\bm{\phi}^{(t)}$, and ${\bm{\varphi}}^{(t)}$, the dual variables $\bm{\mu}_{1}$ and $\bm{\mu}_{2}$ are updated by
\begin{subequations}\label{update_mu}\begin{align}
\bm{\mu}_{1}&:=\bm{\mu}_{1}+\rho(\mathbf{x}^{(t)}-\mathbf{v}^{(t)}),\\
\bm{\mu}_{2}&:=\bm{\mu}_{2}+\rho(\bm{\phi}^{(t)}-\bm{\varphi}^{(t)}).
\end{align}\end{subequations}
\subsection{Summary and Training Strategy}
\label{training}
After alternatively updating the variables as presented in Sec. \ref{ADMM_x}$\sim$\ref{ADMM_mu}, the waveform $\mathbf{x}$ and passive beamforming $\bm{\phi}$ can be obtained through the proposed ADMM-ALM-DFP-based unfolded algorithm by training the learnable parameters $\bm{\kappa}, \bm{\eta}, \tau$, and $\zeta$.
In particular, the learnable parameter set can be defined as $\mathcal{A}=\{\bm{\eta}^{(t,j)},\bm{\kappa}^{(t,j)}, {\tau}^{(t,j,r)}, {\zeta}^{(t,j,r)}|t=1,\ldots,T_{\rm d},~j=1,\ldots,J_{\rm d}$, $r=1, \ldots,R_{\rm d}\}$, where $T_{\rm d}$, $J_{\rm d}$, and $R_{\rm d}$ denote the number of layers for the unfolded ADMM, ALM, and DFP algorithms, respectively. To initialize the model-driven learning framework, we define $\mathbf{B}^{(0)}=\mathbf{D}^{(0)}=\mathbf{I}$, $\bm{\beta}^{(0)}=0$. For all parameter elements
in set $\mathcal{A}$, the DUN is constructed using a multilayer perceptron (MLP) model. Specifically, for each MLP we deploy single hidden layer containing 10 neurons and use the Xavier approach to realize network initialization.

In order to train the DUNs to update learnable parameter set $\mathcal{A}$, the unfolding networks for model-driven learning framework are employed as shown in Fig. \ref{Fig:framework}. Meanwhile, we define the training dataset as $\mathcal{S}=\left\{\mathcal{S}_{1},\mathcal{S}_{2},\ldots,\mathcal{S}_{B}\right\}$, where $\mathcal{S}_{b}$ is the RIS-ISAC system channel state information (CSI) set for the $b$-th training batch. For each CSI set, the forward propagation through DUNs yields the transmit waveform and RIS passive beamforming design, which is actually a mapping operation described by $\left\{\mathbf{x}, \bm{\phi}\right\}=\mathcal{P}(\mathcal{S},\mathbf{x}^{(0)},\bm{\phi}^{(0)},\mathcal{A})$, where the initialization of $\mathbf{x}^{(0)}$ and $\bm{\phi}^{(0)}$ is obtained using the same strategy in \cite{LRJSTSP2022}. Accordingly, the loss function during DUN training can be expressed by
\begin{align}
{L}oss_{1}(\mathcal{S}) = f_{1}(\mathcal{P}(\mathcal{S},\mathbf{x}^{(0)},\bm{\phi}^{(0)},\mathcal{A})).
\end{align}

The proposed model-driven learning algorithm for target detection task is summarized in Algorithm \ref{alg:1}.
It is worth noting that, without the support for complex operations in deep learning libraries, the entire unfolding framework actually performs in terms of real-valued functions, which are obtained through appropriate matrix transformations. Due to space limitations, the operations of matrix transformation and differentiation of real-valued functions are omitted.

\begin{algorithm}[t]
\caption{Model-driven learning algorithm for target detection.}
\label{alg:1}
    \begin{algorithmic}[1]
    \REQUIRE $\mathcal{S}, \mathbf{x}^{(0)}, \bm{\phi}^{(0)}, \bm{\beta}^{(0)}, \mathcal{A}$.
    \ENSURE $\mathbf{x}^{\star}, \bm{\phi}^{\star}$.
       \FOR {$t=1 : T_{\rm d}$}
            \FOR {$j =1: J_{\rm d}$}
                \STATE {Update $\bm{\beta}^{(t,j)}$ by (\ref{update_beta_x}).}
                \FOR{$r=1 : R_{\rm d}$}
                    \STATE  {Update $\mathbf{x}^{(t,j,r)}$ by (\ref{update_x})} and (\ref{x:DFP}).
                \ENDFOR
            \ENDFOR
            \STATE {Update $\mathbf{v}^{(t)}$ by (\ref{update_v}).}
             \FOR {$j=1 : J_{\rm d}$}
                \STATE {Update $\bm{\beta}^{(t,j)}$ by (\ref{update_beta_phi}).}
                \FOR{$r=1 : R_{\rm d}$}
                    \STATE  {Update $\bm{\phi}^{(t,j,r)}$ by (\ref{update_phi})} and (\ref{phi:DFP}).
                \ENDFOR
            \ENDFOR
            \STATE {Update $\bm{\varphi}^{(t)}$ by (\ref{update_varphi}).}
            \STATE {Update $\bm{\mu}^{(t)}$ by (\ref{update_mu}).}
        \ENDFOR
    \end{algorithmic}
\end{algorithm}
    \vspace{-0.2cm}

\section{Joint Waveform and Passive Beamforming Design for DoA Estimation}
Parameter estimation is another important task for radar sensing. In this section, we focus on the DoA estimation of $\bm{\theta}=[{\theta}_{0}, {\theta}_{\rm RIS}]^T$ and propose the joint dual-functional waveform and passive beamforming design to achieve high-accuracy DoA estimation, while simultaneously meeting the required communication QoS. We first define the unknown  parameters of DoA and RCS of the target as $\bm{\varepsilon}\triangleq{[{\bm{\theta}^T,{\bm{\alpha}}_{0}^{T}}]}^{T}$ with ${\bm{\alpha}}_{0}\triangleq{[\Re\{{\alpha}_{0}\},\Im\{{\alpha}_{0}\}]}^{T}$. We recall that  the stacked complex observation  $\mathbf{r}\triangleq[\mathbf{r}[1]^T,\mathbf{r}[2]^T,\ldots,\mathbf{r}[L]^T]^T$ exhibits complex multivariate Gaussian distribution \cite{E1} as $\mathbf{r}\thicksim\mathcal{CN}(\bm{\mu},\mathbf{C})$, where $\bm{\mu}\triangleq{\alpha}_{0}\widetilde{\mathbf{H}}_{0}(\bm{\phi})\mathbf{x}$, $\mathbf{C}\triangleq\sum_{q=1}^{Q}{\xi}_{q}^{2}\widetilde{\mathbf{H}}_{q}(\bm{\phi})\mathbf{x}\mathbf{x}^{H}\widetilde{\mathbf{H}}_{q}^{H}(\bm{\phi})+{\xi}_{\rm z}^{2}\mathbf{I}_{ML}$.
Accordingly, the $(i,j)$-th element of the Fisher information matrix can be calculated as
\begin{align}
\mathbf{F}_{\rm IM}(i,j)&=\mathbb{E}\left[\ddfrac{\partial\log{p(\mathbf{r}|\bm{\varepsilon})}}{\partial{\varepsilon}_{i}}\ddfrac{\partial\log{p(\mathbf{r}|\bm{\varepsilon})}}{{\varepsilon}_{j}}\right]\nonumber\\
\label{Fisher1}
&=\textrm{Tr}\bigg\{\mathbf{C}^{-1}\frac{\partial\mathbf{C}}{\partial{\varepsilon}_{i}}\mathbf{C}^{-1}\frac{\partial\mathbf{C}}{\partial{\varepsilon}_{j}}\bigg\}+2\Re\bigg\{\frac{\partial\bm{\mu}^{H}}{\partial{\varepsilon}_{i}}\mathbf{C}^{-1}\frac{\partial\bm{\mu}}{\partial{\varepsilon}_{j}}\bigg\}.
\end{align}
In order to obtain a closed-form expression of the CRB for DoA estimation, the block form of CRB matrix $\mathbf{E}$ can be defined as
\begin{align}
\mathbf{E}={\begin{bmatrix}
\mathbf{E}_{\bm{\theta}\bm{\theta}^T} & \mathbf{E}_{\bm{\theta}\bm{\alpha}_0^T}\\
\mathbf{E}_{\bm{\theta}\bm{\alpha}_0^T}^T & \mathbf{E}_{\bm{{\alpha}_{0}}\bm{\alpha}_0^T}
\end{bmatrix}}
=\mathbf{F}_{\rm IM}^{-1}
={\begin{bmatrix}
 \mathbf{F}_{\bm{\theta}\bm{\theta}^T} & \mathbf{F}_{\bm{\theta}{\bm{\alpha}}_{0}^{T}}\\
 \mathbf{F}_{\bm{\theta}\bm{\alpha}_0^T}^T & \mathbf{F}_{\bm{\alpha}_{0}\bm{\alpha}_0^T}
\end{bmatrix}}^{-1}.
\end{align}
Thus, the CRB of estimating $\bm{\theta}$ can be expressed as
\begin{align}
\rm{CRB}_{\bm{\theta}}&=\textrm{Tr}\{\mathbf{E}_{\bm{\theta}\bm{\theta}^T}\}\nonumber\\
&=\textrm{Tr}\{(\mathbf{F}_{\bm{\theta}\bm{\theta}^T}-\mathbf{F}_{\bm{\theta}\bm{\alpha}_0^T}\mathbf{F}_{\bm{\alpha}_0\bm{\alpha}_0^T}^{-1}\mathbf{F}_{\bm{\theta}\bm{\alpha}_0^T}^T)^{-1}\}. \label{CRB_define}
\end{align}
The derivations and expressions of $\mathbf{F}_{\bm{\theta}\bm{\theta}^T}$, $\mathbf{F}_{\bm{\theta}\bm{\alpha}_0^T}$, and $\mathbf{F}_{\bm{\alpha}_0\bm{\alpha}_0^T}$ are offered in Appendix \ref{AP1}.

Based on the expression in (\ref{CRB_define}), our goal is to jointly design dual-functional waveform and passive beamforming to minimize the CRB of DoA estimation while satisfying the multi-user communication QoS and the transmit power limitation. Accordingly, this DoA estimation-oriented optimization problem can be formulated as
\begin{subequations}
\label{problem_2}
\begin{align}
&\min_{\mathbf{x},\bm{\phi}}~\textrm{Tr}\{(\mathbf{F}_{\bm{\theta}\bm{\theta}^T}-\mathbf{F}_{\bm{\theta}\bm{\alpha}_0^T}\mathbf{F}_{\bm{\alpha}_0\bm{\alpha}_0^T}^{-1}\mathbf{F}_{\bm{\theta}\bm{\alpha}_0^T}^T)^{-1}\}\label{objective_2}\\
&~~\text{s.t.}~~\rm{(\ref{C:QoS})-(\ref{P:phi})}.
\end{align}
\end{subequations}

It is obvious that problem (\ref{problem_2}) shows its non-convex property, which hinders the application of traditional optimization methods. Moreover, compared to the target detection-oriented problem (\ref{problem}) solved in the previous section, the highly nonlinear nature of the objective function (\ref{objective_2}) leads to poor numerical stability and convergence if using the proposed unfolded ADMM-ALM-DFP method.
To tackle these difficulties, a robust ADMM-PHR-ALM-DFP-based unfolding algorithm is adopted to solve the problem (\ref{problem_2}), which is presented below.

\subsection{Model-driven Learning Framework}
Similar to the ADMM-based framework for the target detection task, after introducing auxiliary variables $\mathbf{v}\triangleq[v_1,v_2,\ldots,v_{ML}]^T$ and $\bm{\varphi}\triangleq[{\varphi}_{1},\varphi_2,\ldots,{\varphi}_{N}]^T$, the DoA estimation-oriented  problem can be transformed to an augmented Lagrangian problem as
\begin{equation}\begin{aligned}\label{ALM_form}
&\min_{\mathbf{x},\bm{\phi},\mathbf{v},\bm{\varphi}}f_{2}(\mathbf{x},\bm{\phi})+\frac{\rho}{2}{\|\mathbf{x}-\mathbf{v}+\frac{\bm{\mu}_{1}}{\rho}\|}^{2}\\
&\qquad\qquad+\frac{\rho}{2}{\|\bm{\phi}-\bm{\varphi}+\frac{\bm{\mu}_{2}}{\rho}\|}^{2}+\mathbb{I}(\mathbf{x}, \bm{\phi},\mathbf{v},\bm{\varphi}),
\end{aligned}\end{equation}
where  we define the CRB objective function $f_{2}(\mathbf{x},\bm{\phi})=\textrm{Tr}\{(\mathbf{F}_{\bm{\theta}\bm{\theta}^T}-\mathbf{F}_{\bm{\theta}\bm{\alpha}_0^T}\mathbf{F}_{\bm{\alpha}_0\bm{\alpha}_0^T}^{-1}\mathbf{F}_{\bm{\theta}\bm{\alpha}_0^T}^T)^{-1}\}$ for brief expression, $\rho>0$ is the penalty parameter, $\bm{\mu}_{1}$, $\bm{\mu}_{2}$ denote the ADMM dual variables, and $\mathbb{I}(\mathbf{x}, \bm{\phi},\mathbf{v},\bm{\varphi})$ is the indicator function to impose constraints (\ref{C:QoS2})-(\ref{C2}), (\ref{C:v}), and (\ref{C:varphi}).

Then, the augmented Lagrangian problem (\ref{ALM_form}) can be decomposed into multiple sub-problems by alternately updating variables $\mathbf{x},\bm{\phi},\mathbf{v}, \bm{\varphi}$, and $\bm{\mu}$. By adopting the similar ADMM-based framework, the updates of $\mathbf{v}, \bm{\varphi}$, $\bm{\mu}_1$, and $\bm{\mu}_2$ use the same operations as (\ref{update_v}), (\ref{update_varphi}), and (\ref{update_mu}), respectively.
However, it is worth emphasizing that the objective function (\ref{objective_2}) exhibits strong nonlinearity and non-convexity. Consequently, directly applying the Lagrangian multiplier method to handle those linear inequality constraints (21b) may lead to severe convergence challenges. To address this issue, we propose to employ the PHR-based transformation to reformulate these linear constraints into a nonlinear form. This approach is expected to better align with the nonlinear and non-convex properties of the objective function, thereby enhancing the convergence and feasibility of the solution.

\subsection{Update $\mathbf{x}$}
In the $t$-th ADMM iteration, with given $\mathbf{v}^{(t)}$, $\bm{\phi}^{(t)}$, $\bm{\varphi}^{(t)}$, ${\bm{\mu}}_{1}$ and ${\bm{\mu}}_{2}$, the ALM problem with respect to variable $\mathbf{x}^{(t)}$ can be formulated as
\begin{align}
\label{x_initial}
\min_{\mathbf{x}}~&f_{2}(\mathbf{x},\bm{\phi}^{(t)})
+\frac{\rho}{2}\|\mathbf{x}-\mathbf{v}^{(t)}+\frac{\bm{\mu}_{1}}{\rho}\|^2+\mathbb{I}(\mathbf{x}, \bm{\phi}^{(t)},\mathbf{v}^{(t)},\bm{\varphi}^{(t)}).
\end{align}
Then, we employ a modified PHR transformation to convert the linear inequality constraints (\ref{C:QoS}) inherent in the indicator function $\mathbb{I}(\mathbf{x}, \bm{\phi}^{(t)},\mathbf{v}^{(t)},\bm{\varphi}^{(t)})$ to a nonlinear form. In specific, by introducing the slack variable $ p_{k,l}\geq0$, $\forall k=1,\ldots, M$, $\forall l=1,\ldots, L$, the inequality constraint $g_{k,l}(\mathbf{x},\bm{\phi})\leq 0$ can be converted into equality constraint $g_{k,l}(\mathbf{x},\bm{\phi})+p_{k,l} = 0$, and then the augmented Lagrangian problem is formulated as
\begin{align}
    \min_{\mathbf{x},\bm{\beta}}~\min_{\mathbf{p}}~~&f_{2}(\mathbf{x},\bm{\phi}^{(t)})+\frac{\rho}{2}\|\mathbf{x}-\mathbf{v}^{(t)}+\frac{\bm{\mu}_{1}}{\rho}\|^2\\
    &+\sum_{l=1}^{L}\sum_{k=1}^{K}\frac{{\eta}_{k,l}}{2}\big\|g_{k,l}(\mathbf{x},\bm{\phi}^{(t)})+{p_{k, l}}+\frac{{\beta}_{k, l}}{{\eta}_{k, l}}\big\|^2\label{subxs},\nonumber
\end{align}
where $\eta_{k,l}$, $k=1,\ldots, K$, $l=1,\ldots,L$ denote the learnable step-size parameter for the PHR-ALM dual update, $\beta_{k,l}$, $k=1,\ldots, K$, $l=1,\ldots, L$, represent the Lagrangian multiplier. Similarly, we also define the vector $\bm{\beta}$ that collects all Lagrangian multipliers as $\bm{\beta}\triangleq[\beta_{1,1},  \ldots, \beta_{K,L}]^T\in\mathbb{R}^{KL}$ for clearer notation.
According to the derivations in \cite{ALM}, we can obtain the closed-form solution to $p_{k,l}$ as $p_{k,l}^\star = \max\{-g_{k, l}(\mathbf{x},\bm{\phi})-\beta_{k,l}/\eta_{k,l},0\}$. Substituting $p_{k,l}^\star $ into (45), the augmented Lagrangian problem can be converted into
\be\begin{aligned}
\underset{\mathbf{x},\bm{\beta}}\min~~&\mathcal{L}_{2}(\mathbf{x},\bm{\beta})=f_{2}(\mathbf{x},\bm{\phi}^{(t)})+\frac{\rho}{2}{\|\mathbf{x}-\mathbf{v}^{(t)}+\frac{\bm{\mu}_{1}}{\rho}\|^2}\\    &+\sum_{l=1}^{L}\sum_{k=1}^{K}\frac{{\eta}_{k,l}}{2}{\big\|\max\{g_{k,l}(\mathbf{x},\bm{\phi}^{(t)})+\frac{{\beta}_{k,l}}{{\eta}_{k,l}}, 0\}\big\|^2}.\label{sub_1_CRB}
\end{aligned}\ee
Thus, in the $j$-th PHR-ALM iteration, the update for $\beta^{(t,j)}_{k,l}$ and $\mathbf{x}^{(t,j)}$ are given by
\begin{subequations}\begin{align}
    &\beta_{k,l}^{(t,j)}=\max\{{\beta}^{(t,j-1)}_{k,l}+{\eta}_{k,l}^{(t,j)}g_{k,l}(\mathbf{x}^{(t,j-1)},\bm{\phi}^{(t)}), 0\}\label{update_beta_x_2}.\\
    &\mathbf{x}^{(t,j)}=\arg\min_\mathbf{x}~\mathcal{L}_{2}(\mathbf{x}, \bm{\beta}^{(t,j)})\label{PHR_ALM_x}.
\end{align}\end{subequations}

In the update for $\mathbf{x}^{(t,j)}$, the non-convexity of $\mathcal{L}_{2}(\mathbf{x},\bm{\beta})$ is the major difficulty. To address this issue, we employ the unfolding-friendly DFP algorithm developed in Sec. \ref{ADMM_x} to further decompose problem (\ref{PHR_ALM_x}) into a series of closed-form solutions.
Particularly, the update of $\mathbf{x}^{(t,j)}$ in the $r$-th DFP iteration is expressed as
\begin{align}
\mathbf{x}^{(t,j,r)}=\mathbf{x}^{(t,j,r-1)}-{\tau}^{(t,j,r)}\mathbf{B}^{(r-1)}\nabla\mathcal{L}_{2}(\mathbf{x}^{(t,j,r-1)}) \label{update_x_2},
\end{align}
where ${\tau}^{(t,j,r)}$ is the learnable step-size parameter for DFP algorithm. The matrix $\mathbf{B}^{(r-1)}$ is updated following a procedure similar to that in (\ref{SPDM}). To ensure that $\mathbf{x}$ satisfies the power constraint (\ref{C1}), the vector $\mathbf{x}^{(t,j,r)}$ is then scaled into the feasible set as
\begin{align}
\label{x:DFP2}
\mathbf{x}^{(t,j,r)}(i):=\frac{\sqrt{P/M}\mathbf{x}^{(t,j,r)}(i)}{|\mathbf{x}^{(t,j,r)}(i)|},~\forall i=1,2,\ldots,ML.
\end{align}

\subsection{Update $\bm{\phi}$}
With given $\mathbf{x}^{(t)}$, $\mathbf{v}^{(t)}$, $\bm{\varphi}^{(t)}$, ${\bm{\mu}}_{1}$ and ${\bm{\mu}}_{2}$, the objective problem with respect to variable $\bm{\phi}$ can be formulated as
\begin{align}
\min_{\bm{\phi}}~&f_{2}(\bm{\phi},\mathbf{x}^{(t)})
+\frac{\rho}{2}\|\bm{\phi}-\bm{\varphi}+\frac{\bm{\mu}_{2}}{\rho}\|^2+\mathbb{I}(\bm{\phi},\mathbf{x}^{(t)},\mathbf{v}^{(t)},\bm{\varphi}^{(t)}).
\end{align}
Similarly, by utilizing PHR-ALM, the Lagrangian function with respect to variables $\bm{\phi}^{(t)}$ and the Lagrange multiplier $\bm{\beta}^{(t)}$ is given as
\begin{align}
    \mathcal{L}_{2}(\bm{\phi},\bm{\beta})&=f_{2}(\bm{\phi},\mathbf{x}^{(t)})+\frac{\rho}{2}\|\bm{\phi}-\bm{\varphi}^{(t)}+\frac{\bm{\mu}_{2}}{\rho}\|^2 \\
&+\sum_{l=1}^{L}\sum_{k=1}^{K}\frac{{\kappa}_{k,l}}{2}\big\|\max\{g_{k,l}(\bm{\phi},\mathbf{x}^{(t)})+\frac{\beta_{k,l}}{{\kappa}_{k,l}}, 0\}\big\|^2, \nonumber
\end{align}
where $\kappa_{k,l}$, $k=1,\ldots,K$, $l=1,\ldots,L$ denote the learnable step-size parameter for the PHR-ALM dual update.
In the $j$-th PHR-ALM iteration, $\bm{\phi}^{(t,j)}$ and $\beta^{(t,j)}_{k,l}$ are determined as follows
\begin{subequations}\begin{align}
\beta^{(t,j)}_{k,l}&=\max\left\{\beta^{(t,j-1)}_{k,l}+{\kappa}^{(t,j)}_{k,l}g_{k,l}(\bm{\phi}^{(t,j-1)},\mathbf{x}^{(t)}), 0\right\}, \label{update_beta_phi_2}\\
\bm{\phi}^{(t,j)}&=\arg\min_{\bm{\phi}}~\mathcal{L}_{2}(\bm{\phi}, \bm{\beta}^{(t,j)}).\label{sub_2_CRB}
\end{align}\end{subequations}

After using the PHR-ALM to cope with inequality constraints,  we employ a modified unfolding-friendly DFP algorithm to solve the non-convex unconstrained optimization sub-problem (\ref{sub_2_CRB}). Specifically, the $r$-th iteration of the generalized Quasi-Newtonian method can be formulated as
\begin{align}
\bm{\phi}^{(t,j,r)}=\bm{\phi}^{(t,j,r-1)}-{{\zeta}}^{(t,j,r)}\mathbf{D}^{(r-1)}\nabla\mathcal{L}_{2}(\bm{\phi}^{(t,j,r-1)})\label{update_phi_2},
\end{align}
where  ${\zeta}^{(t,j,r)}$ denotes the learnable step-size parameter for DFP algorithm. The approximate Hessian matrix $\mathbf{D}$   can be obtained and updated similar to (\ref{SPDM}). To ensure that $\bm{\phi}$ satisfies the constraint (\ref{C2}), $\bm{\phi}^{(t,j,r)}$ needs to be scaled into the feasible set, which can be calculated by
\begin{align}
\label{phi:DFP2}
\bm{\phi}^{(t,j,r)}(n):=\frac{{\bm{\phi}^{(t,j,r)}}(n)}{|\bm{\phi}^{(t,j,r)}(n)|},~\forall n = 1,2,\ldots,N.
\end{align}

\subsection{Summary and Training Strategy}
After alternatively updating the variables $\mathbf{x}$, $\mathbf{v}$, $\bm{\phi}$, $\bm{\varphi}$, $\bm{\mu}_{1}$, and $\bm{\mu}_{2}$, the dual-functional waveform and RIS passive beamforming can be optimized through the ADMM-PHR-ALM-DFP-based model-driven learning framework.
Similarly, the learnable parameter set is defined as $\mathcal{A}=\{\bm{\eta}^{(t,j)},\bm{\kappa}^{(t,j)}, {\tau}^{(t,j,r)}, {\zeta}^{(t,j,r)}|t=1,\ldots,T_{\rm e},~j=1,\ldots,J_{\rm e}$,~$r=1, \ldots,R_{\rm e}\}$, where $T_{\rm e}$, $J_{\rm e}$, and $R_{\rm e}$ denote the number of unfolded-ADMM layers, unfolded-PHR-ALM layers, and unfolded DFP layers, respectively. In order to initialize this model-driven learning framework, it is assumed that $\mathbf{B}^{(0)}=\mathbf{D}^{(0)}=\mathbf{I}$, $\bm{\beta}^{(0)}=0$. For all parameter elements
in set $\mathcal{A}$, the DUN is constructed by deploying an MLP model. In particular, each hidden layer in DUNs contains 10 neurons. All DUNs adopt the well-known Xavier approach to realize parameter initialization.

By inputting the standard CSI set $\mathcal{S}$ mentioned before, the loss function during DUNs training can be accordingly expressed by
\begin{align}
{L}oss_{2}(\mathcal{S}) = f_{2}(\mathcal{P}(\mathcal{S},\mathbf{x}^{(0)},\bm{\phi}^{(0)},\mathcal{A})).
\end{align}

After DoA estimation-oriented training, the dual-functional waveform $\mathbf{x}$ and passive RIS coefficient $\bm{\phi}$ are well-designed, and the detailed model-driven learning algorithm for DoA estimation task is accordingly shown in Algorithm \ref{alg:2}.
\begin{algorithm}[t]
\caption{Model-driven learning algorithm for DoA estimation.}
\label{alg:2}
    \begin{algorithmic}[1]
    \REQUIRE $\mathcal{S}, \mathbf{x}^{(0)}, \rm{\phi}^{(0)}, \bm{\beta}^{(0)}, \mathcal{A}$.
    \ENSURE $\mathbf{x}^{\star}, \bm{\phi}^{\star}$.
       \FOR {$t=1:T_{\rm e}$}
            \FOR {$j=1 : J_{\rm e}$}
                \STATE {Update $\bm{\beta}^{(t,j)}$ by (\ref{update_beta_x_2})}
                \FOR{$r=1 : R_{\rm e}$}
                    \STATE  {Update $\mathbf{x}^{(t,j,r)}$ by (\ref{update_x_2}) and (\ref{x:DFP2})}
                \ENDFOR
            \ENDFOR
            \STATE {Update $\mathbf{v}^{(t)}$ by (\ref{update_v})}
             \FOR {$j=1 : J_{\rm e}$}
                \STATE {Update $\bm{\beta}^{(t,j)}$ by (\ref{update_beta_phi_2})}
                \FOR{$r=1 : R_{\rm e}$}
                    \STATE  {Update $\bm{\phi}^{(t,j,r)}$ by (\ref{update_phi_2}) and (\ref{phi:DFP2})}
                \ENDFOR
            \ENDFOR
            \STATE {Update $\bm{\varphi}^{(t)}$ by (\ref{update_varphi})}
            \STATE {Update $\bm{\mu}^{(t)}$ by (\ref{update_mu})}
        \ENDFOR
    \end{algorithmic}
\end{algorithm}
\section{Simulation Results}
\label{Sim}
In this section, we provide extensive simulation results to validate the performance of our proposed model-driven learning algorithm for joint waveform and beamforming designs. The following settings are used throughout the paper unless specified: $M = 6$, $N=64$, QPSK-modulated symbols, $K=3$, $L=20$, $P=20$dBW, ${\xi}_{0}^{2}={\xi}_{q}^{2}=1$, ${\xi}_{\rm z}^{2}={\sigma}^2_{k}=-80$dBm, ${\Gamma}_{k}=10$dB, $\forall k$. The range-angle location of the target is $(0,0^{\circ})$. Meanwhile, there are $Q=3$ clutters at locations $(0,{-60}^{\circ})$, $(1,{-20}^{\circ})$, and $(2,{45}^{\circ})$, respectively. Typical path-loss model $\rm{P}(\it d)=C_{0}{(d_{0}/d)}^{\varrho},$ is employed in this paper, where $C_{0}$ = -30dB, $d_{0}$ =1m, $d$ represents the distance of link, and $\varrho$ denotes the path-loss exponent that generally varies from 2 to 4. Accordingly, the path loss exponents for the BS-RIS, RIS-target, RIS-user, BS-target, and BS-user transmission links are set to 2.3, 2.3, 2.5, 2.8, and 3, respectively. During model-driven training, each ADMM framework is trained in a layer-by-layer basis. The size of the dataset is $B=800$. All parameters mentioned above are fixed without special declaration. The number of different trainable parameters in target detection task are set as $T_{\rm d}=75$, $J_{\rm d}=8$, and $R_{\rm d}=10$, while the number of different trainable parameters in parameter detection task are set as $T_{\rm e}=60$, $J_{\rm e}=10$, and $R_{\rm e}=10$, respectively.

\subsection{Target Detection Performance}
We first evaluate the radar output SINR performance in the target detection task and analyze the quality of the designed transmit dual-functional waveforms. To demonstrate the target detection performance of our proposed model-driven learning algorithm (denoted as ``\textbf{Prop., w/ RIS}'') and the gains brought by RIS, the performance of following schemes are also evaluated for comparisons.
\begin{itemize}
\item ``\textbf{Prop., rand. RIS}'': Without designing RIS passive beamforming $\bm{\phi}$ in RIS-ISAC systems, only the dual-functional waveform $\mathbf{x}$ is designed by using our proposed model-driven learning algorithm.
\item ``\textbf{Prop., w/o RIS}'': Without deploying RIS in ISAC systems, only the dual-functional waveform $\mathbf{x}$ is solved by using our proposed model-driven learning algorithm.
\item ``\textbf{ADMM-MM}'': This scheme is a convex optimization-based algorithm proposed in \cite{LRJSTSP2022}, which jointly optimizes dual-functional waveform $\mathbf{x}$ and RIS passive beamforming $\bm{\phi}$ by using complex operations of iterative ADMM and MM approaches.
\item ``\textbf{Radar-only}'': Without considering the multi-user communication QoS, the transmitted dual-functional waveform $\mathbf{x}$ and RIS passive beamforming are jointly designed only for target detection task.
\end{itemize}

\begin{figure}[!t]
\centering
\includegraphics[width=3.5in]{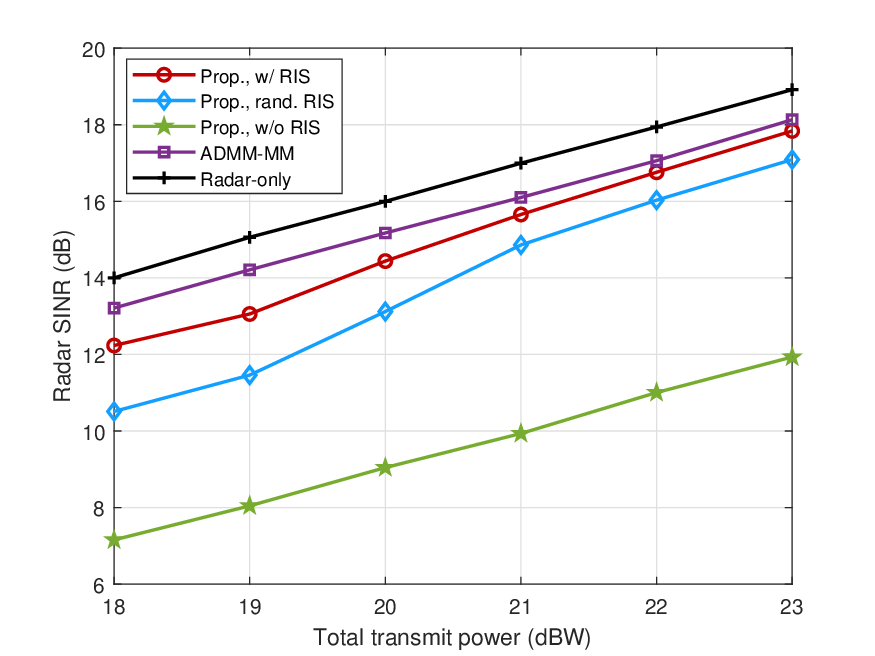}
\caption{Radar SINR versus transmit power $P$ ($N=64$).}
\label{Sim:P}  \vspace{-0.0 cm}
\end{figure}
\begin{figure}[!t]
\centering
\includegraphics[width=3.5in]{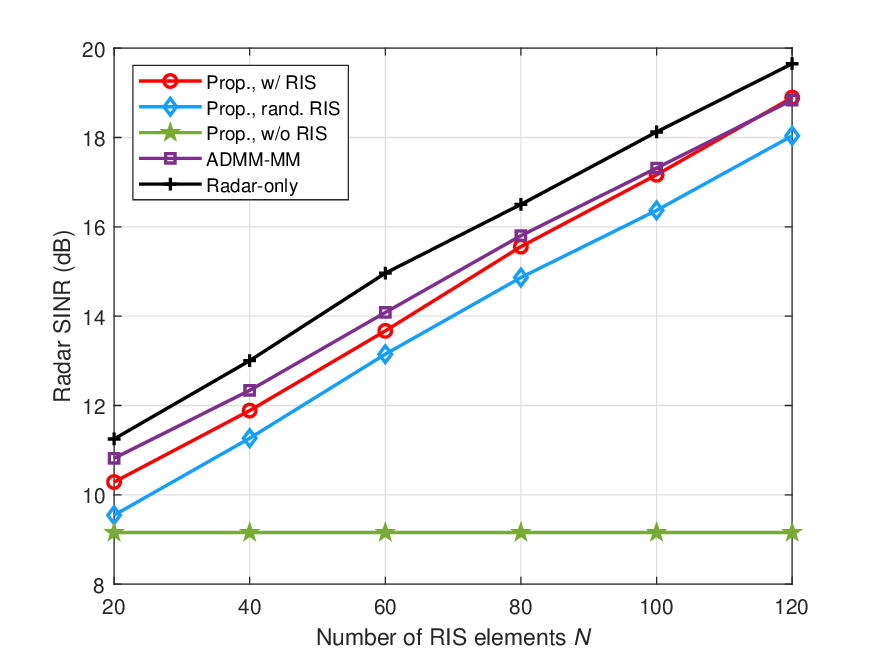}
\caption{Radar SINR versus the number of RIS element $N$.}
\label{Sim:N}
\vspace{0.0cm}
\end{figure}
As shown in Fig. \ref{Sim:P}, the relationship between the radar SINR and transmit power $P$ is plotted. Not surprisingly, the radar SINR increases with the increase of $P$. It is obvious that our proposed model-driven learning-based dual-functional scheme shows low-level radar performance loss to the radar-only scheme. Compared with the scheme without RIS and random RIS coefficient, both the proposed approach and the optimization algorithm in \cite{LRJSTSP2022} achieve much higher radar SINR, i.e., about 6dB radar SINR gains over the scheme without RIS. In addition, we observe that the proposed model-driven learning approach has comparable performance and exhibits marginal performance loss compared to the conventional ADMM-MM scheme. Moreover, the performance gap becomes negligible as $P$ increases.

Then, the radar SINR versus the number of RIS elements $N$ is shown in Fig. \ref{Sim:N}. Similar relationships and conclusions can also be found as that from Fig. \ref{Sim:P}. Our proposed model-driven learning method achieves comparable performance to the radar-only scheme and ADMM-MM algorithm \cite{LRJSTSP2022}. It is also noteworthy that as the number of RIS elements increases, our proposed model-driven learning approach gets closer radar performance to the ADMM-MM algorithm, which will highlight the huge advantages of our proposed model-driven learning algorithm in large-scale RIS deployment scenarios.

\begin{figure}[!t]
\centering
\includegraphics[width=3.5in]{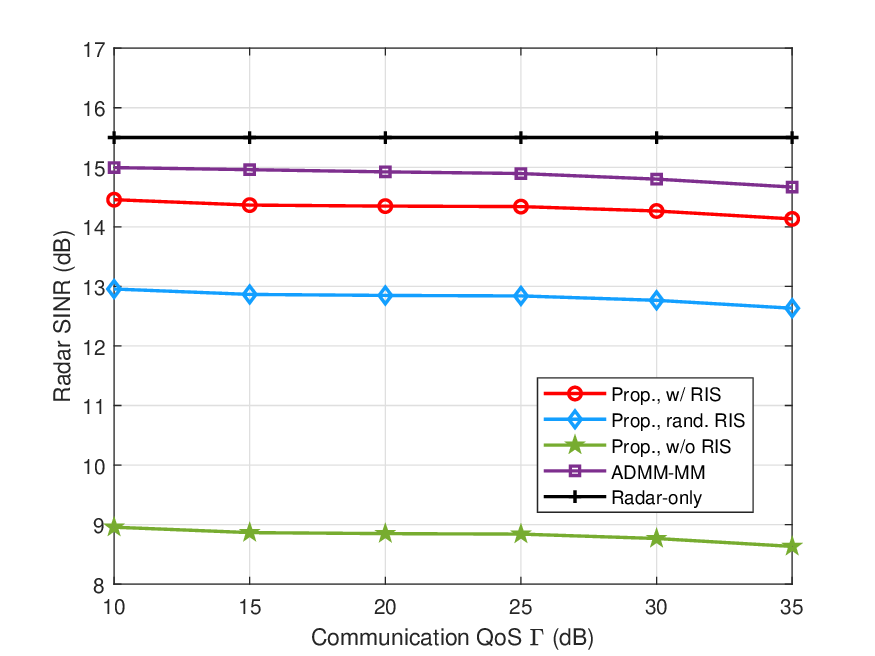}
\caption{Radar SINR Versus multi-user communication QoS $\Gamma$.}
\label{SINR_Gamma}
\vspace{-0.0 cm}
\end{figure}

To illustrate the trade-offs in the target detection task, Fig. \ref{SINR_Gamma} depicts the relationship between the radar output SINR and the multi-user communication QoS $\Gamma$. It can be observed that increasing $\Gamma$ has negligible impact on radar sensing performance when $\Gamma$ remains within a reasonable range. This behavior arises because the primary objective is to maximize the radar output SINR, and the resulting dual-functional waveform inherently supports high QoS under the CI-based metric. However, a trade-off between radar sensing performance and excessively high communication QoS becomes evident as $\Gamma$ continues to increase.

\begin{figure}[!t]
\centering
\includegraphics[width=3.5in]{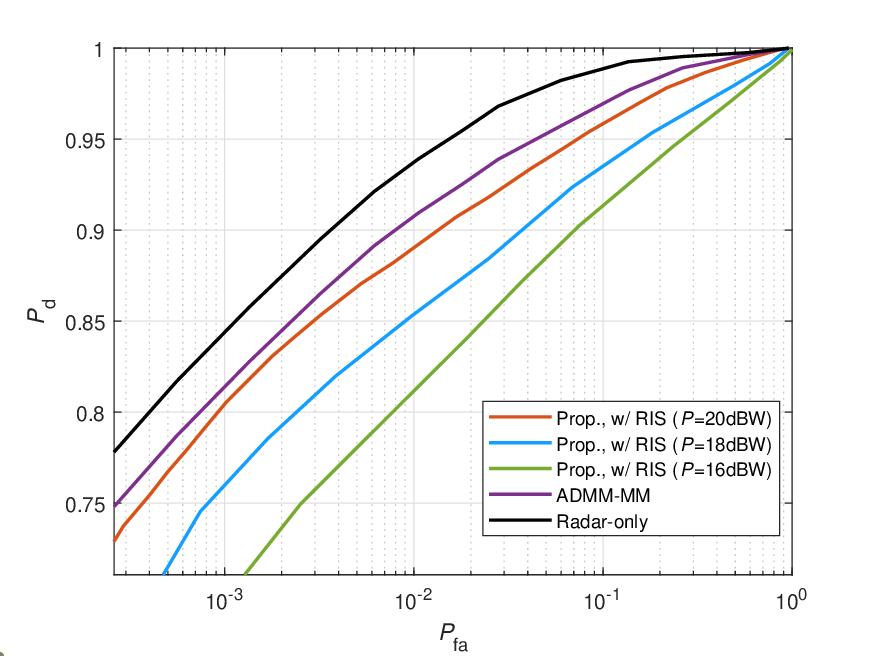}
\caption{Detection probability versus false alarm probability.}
\label{ROC}
\vspace{0.0cm}
\end{figure}

 To more intuitively demonstrate the detection performance, we show the receiver operating characteristic (ROC) in Fig. \ref{ROC}, which provides the relationship between the detection probability $P_{\rm d}$ and the false alarm probability $P_{\rm fa}$. The results clearly indicate that the proposed model-driven learning scheme achieves performance comparable to the ADMM-MM-based method and the radar-only scheme in the target detection task.

The computational complexity is assessed based on the average execution time, as presented in Table \ref{Sim:time_1}. It is evident that the proposed model-driven learning algorithm significantly outperforms the ADMM-MM algorithm in terms of execution time, achieving a reduction of approximately $70\%$ for $N=128$. This substantial reduction in complexity highlights the practical advantages of the proposed approach, particularly given its comparable performance to existing methods, as demonstrated in Figs. \ref{Sim:P} and \ref{Sim:N}. The reduced average execution time underscores the efficiency of the deep learning-based framework, which offers a remarkable improvement over conventional symbol-level techniques with similar radar performance.

\begin{table}[!t]
\centering %
\caption{Average execution time of detection task (second)}
\begin{tabular}{cccc}
\toprule
\multicolumn{1}{m{2.8cm}}{\centering $N$} &\multicolumn{1}{m{1.25cm}}{\centering 32} & \multicolumn{1}{m{1.25cm}}{\centering 64} & \multicolumn{1}{m{1.25cm}}{\centering 128}\\
\midrule %
ADMM-MM&0.414&3.490&22.136\\

Proposed method&0.168&0.976&6.734\\
\bottomrule
\end{tabular}
\label{Sim:time_1}
\end{table}
\subsection{DoA Estimation Performance}
In this subsection, we evaluate the CRB performance for DoA estimation task. In order to demonstrate the DoA estimation performance of our proposed model-driven learning algorithm (denoted as ``\textbf{Prop., w/ RIS}'') and the gains brought by RIS, the following schemes are introduced for comparisons.
\begin{itemize}
\item ``\textbf{Unsuper., w/ RIS}'': This scheme is a traditional unsupervised deep learning approach, which jointly optimizes transmitted dual-functional waveform $\mathbf{x}$ and RIS passive beamforming $\bm{\phi}$ by employing the objective function loss and constraint-based regular terms.
\item ``\textbf{Prop., rand. RIS}'': Without designing RIS passive beamforming $\bm{\phi}$ in RIS-ISAC systems, only the dual-functional waveform $\mathbf{x}$ is designed by using our proposed model-driven learning algorithm.
\item ``\textbf{Radar-only}'': Without considering the multi-user communication QoS, the transmitted dual-functional waveform $\mathbf{x}$ and RIS passive beamforming $\bm{\phi}$ are jointly designed only for DoA estimation task.
\end{itemize}

\begin{figure}[!t]
\centering
\includegraphics[width=3.5in]{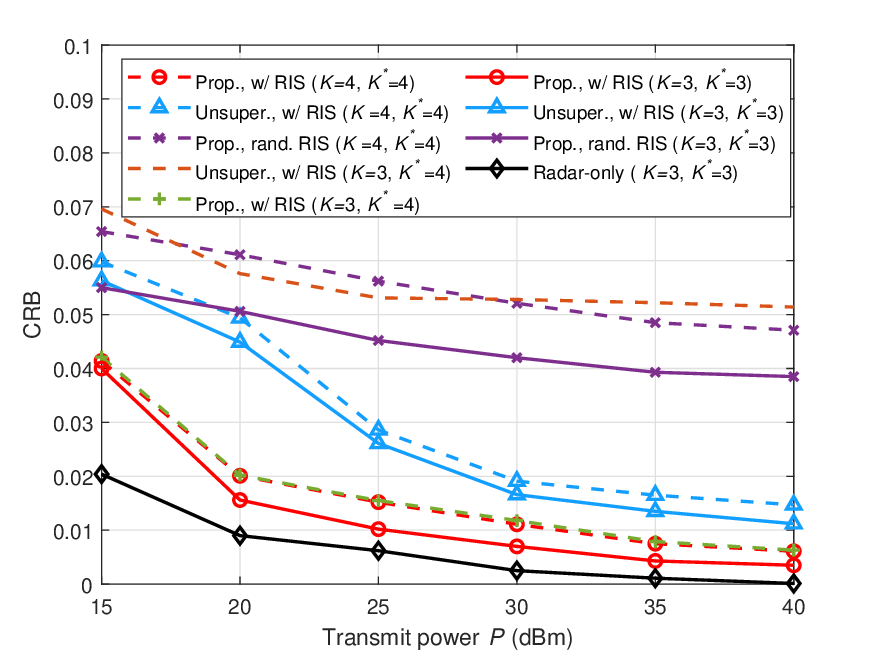}
\caption{DoA estimation CRB versus transmit power $P$ ($N$=64).}
\label{CRB_P}
\vspace{0.4cm}
\end{figure}
In Fig. \ref{CRB_P}, we first illustrates the relationship between the CRB and the transmit power $P$.
Unlike the target detection task, there is no convex optimization based joint waveform and beamforming design for the considered DoA estimation task due to the complicated optimzation problem. Therefore, for comparison purpose, we include result from the radar-only scheme and an unsupervised deep learning method which also exhibits low-complexity. Moreover, the performance of the proposed algorithm for ISAC scenario with random RIS coefficients is also evaluated. As expected, the CRB for DoA estimation decreases as the transmit power increases. The proposed model-driven learning scheme shows slightly loss compared to the radar-only scheme, thereby demonstrating the excellent DoA estimation performance of our proposed method. Notably, our proposed model-driven learning approach demonstrates significantly superior DoA estimation performance compared to both the scenario with random RIS coefficients and the unsupervised learning method. To further validate the generalization ability and robustness of the proposed model-driven learning scheme, we test the performance of the model-driven learning framework with the parameters $\mathcal{A}$ trained at scenarios with different number of communication users, i.e., $K$ users during the training stage and $K^{*}$ served users during the validation process. The results clearly demonstrate that the proposed model-driven scheme achieves comparable performance for the transferred model and the training model, even when the number of communication users changes. In contrast, the unsupervised learning-based scheme exhibits significant CRB performance degradation for the transferred model when the number of served users changes. This indicates that the proposed model-driven learning framework effectively mitigates the generalization challenges faced by deep learning models and ensures robust performance for ISAC system designs. Furthermore, compared to the CRB performance with $K = 3$ users, this observation highlights the inherent trade-off between communication and radar functionalities in ISAC systems.

\begin{figure}[!t]
\centering
\includegraphics[width=3.5in]{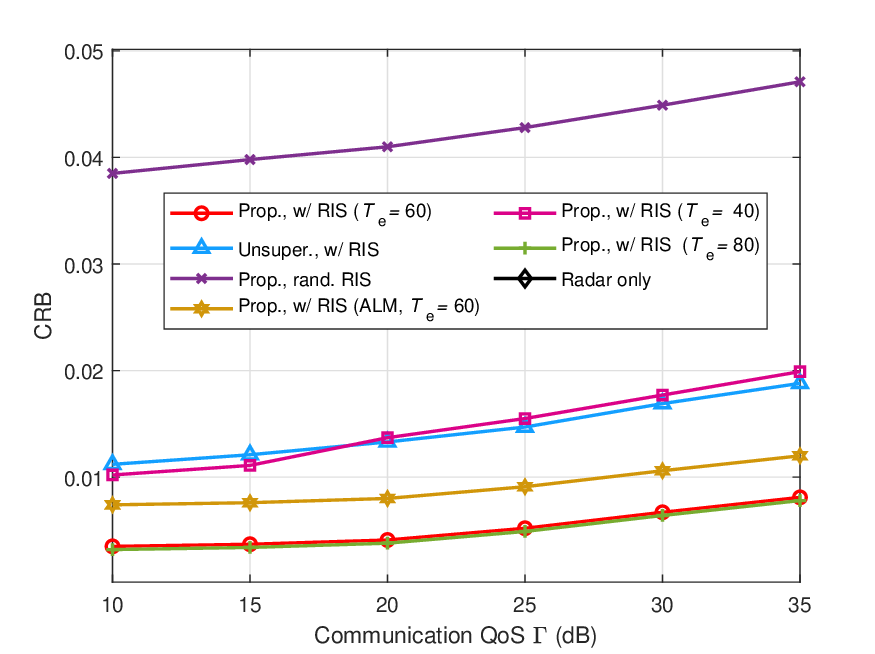}
\caption{DoA estimation CRB versus multi-user communication QoS $\Gamma$ ($P$ = 40dBm).}
\label{CRB_Gamma}
\vspace{0.0cm}
\end{figure}

%


In order to further explore the relationship and performance trade-off between sensing and communication functionalities in this RIS-ISAC system, in Fig. \ref{CRB_Gamma} we demonstrate the DoA estimation CRB versus the multi-user communication QoS $\Gamma$.
As illustrated in Fig. \ref{CRB_Gamma}, the CRB progressively increases with higher communication QoS $\Gamma$, further confirming the trade-off between communication and sensing performance in the ISAC system. Meanwhile,  our proposed methods achieve performance comparable to the radar-only scheme and outperform the unsupervised learning and random RIS coefficients scheme, demonstrating the superiority of our approach. Additionally, the convergence performance of our model-driven learning algorithm is validated by varying the number of outer-loop ADMM iterations. For the same number of iterations, our proposed PHR-ALM-based framework significantly outperforms the ALM-based framework, further confirming its superiority in CRB minimization tasks.
Meanwhile, as the number of outer-loop ADMM increases, the DoA estimation CRB obtained by the proposed model-driven learning method gradually decreases and can finally achieve stable convergence.

The conmplexity analysis based on average execution time of DoA estimation is provided in Table \ref{Sim:time_2}.
It can be found that our proposed model-driven learning algorithm has a comparable average execution time to the unsupervised learning scheme, implying that both of them show the low computational complexity. Fortunately, compared to the unsupervised learning scheme, our proposed method demonstrates great advantages in both sensing and communication performance. Meanwhile, compared with the unsupervised learning scheme, our proposed model-driven learning model has stronger generalization ability, can adapt to the interference caused by communication users, potential targets, and environmental changes, which reduces a large amount of training overhead, thus making it more suitable for practical applications and deployments.
\begin{table}[t]
\centering %
\caption{Average execution time of estimation (second)}
\begin{tabular}{cccc}
\toprule
\multicolumn{1}{m{2.8cm}}{\centering $N$} &\multicolumn{1}{m{1.25cm}}{\centering 32} & \multicolumn{1}{m{1.25cm}}{\centering 64} & \multicolumn{1}{m{1.25cm}}{\centering 128}\\
\midrule %
Proposed method&0.296&1.468&7.991\\

Unsupervised learning&0.398&1.517&5.963\\
\bottomrule
\end{tabular}
\label{Sim:time_2}
\end{table}

\subsection{Multi-user Communication Performance}
\begin{figure}
\centering
\includegraphics[width=3.5in]{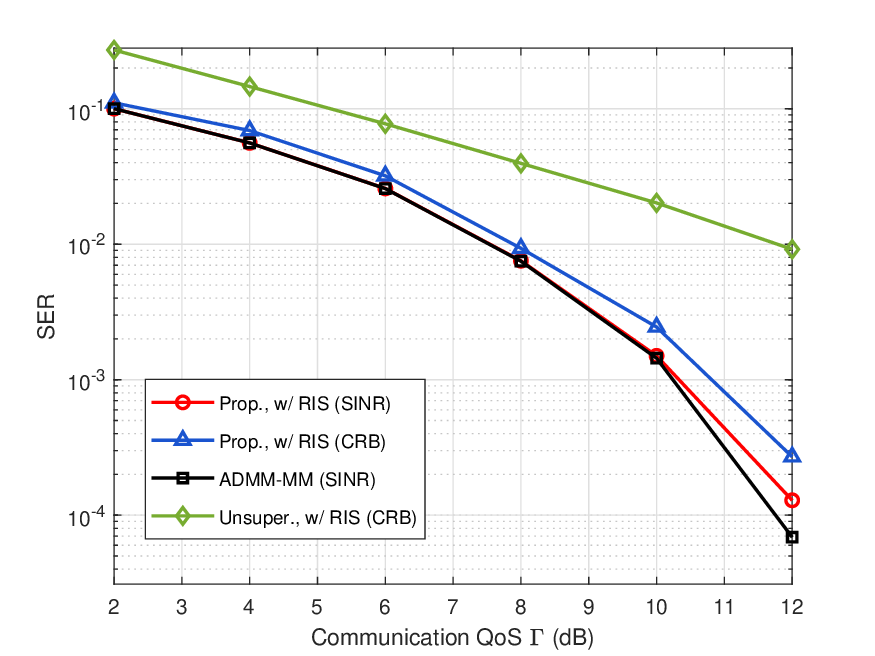}
\caption{SER versus multi-user communication QoS $\Gamma$.}
\label{SER}
\end{figure}

Finally, to thoroughly assess the communication performance of the proposed model-driven learning approach, Fig. \ref{SER} illustrates the relationship between SER and the communication QoS threshold $\Gamma$, across different radar tasks and algorithms. As observed, with increasingly stringent QoS thresholds, the average SER progressively decreases. Notably, the proposed model-driven learning algorithm demonstrates satisfactory multi-user communication performance in both target detection and DoA estimation tasks. In comparison to the unsupervised learning scheme, the proposed model-driven learning algorithm achieves significantly lower SER, particularly excelling in the DoA estimation task. Additionally, the proposed model-driven learning algorithm  exhibits comparable communication performance to the highly complex ADMM-MM method in the target detection task.

\vspace{0.4cm}
\section{Conclusions}
In this paper, we considered symbol-level precoding based joint waveform and passive beamforming design in RIS-aided ISAC systems. We thoroughly investigate problems of radar target detection and direction-of-arrival (DoA) estimation within the ISAC systems. Efficient model-driven learning algorithms were proposed to solve the challenging sensing-oriented ISAC tasks and reduce the computational complexity with comparable performance.
Simulation results demonstrate the superiority of the proposed algorithms in maintaining satisfactory performance for both target detection and DoA estimation, while also ensuring excellent communication quality-of-service (QoS). Additionally, complexity analysis based on average execution time highlights the low computational burden of our approach. This study underscores the potential of model-driven learning strategies to enable low-complexity, high-reliability, and highly generalizable waveform designs for ISAC systems.

\appendices
\section{}
\label{AP1}
In Appendix \ref{AP1}, we provide the brief derivations and expressions of the sub-matrices of Fisher information matrix $\mathbf{F}_{\bm{\theta}\bm{\theta}^T}$, $\mathbf{F}_{\bm{\theta}\bm{\alpha}_0^T}$, and $\mathbf{F}_{\bm{\alpha}_0\bm{\alpha}_0^T}$, respectively.
First, the derivatives of $\bm{\mu}$ with respect to the unknown parameters can be respectively given according to (\ref{Fisher1}) as
\begin{subequations}
    \begin{align}
        \frac{\partial \bm{\mu}}{\partial\theta_{1}}=\alpha_{0}\frac{\partial{\widetilde{\mathbf{H}}_{0}(\bm{\phi})}}{\partial\theta_{0}}\mathbf{x},\\
        \frac{\partial \bm{\mu}}{\partial\theta_{2}}=\alpha_{0}\frac{\partial\widetilde{\mathbf{H}}_{0}(\bm{\phi})}{\partial\theta_{\rm RIS}}\mathbf{x},\\
        \frac{\partial \bm{\mu}}{\partial\alpha_{0}}={[1~ \jmath]}^{T}\otimes \widetilde{\mathbf{H}}_{0}(\bm{\phi})\mathbf{x}.
    \end{align}
\end{subequations}
Notice that $\mathbf{C}$ is independent of both parameters $\bm{\theta}$ and $\alpha_0$, the elements of Fisher information matrix can be accordingly written as
\begin{subequations}
    \begin{align}
        {F}_{{\theta}_{0},{\theta}_{0}}&=2{\xi}_{0}^{2}\Re\{\mathbf{x}^{H}\frac{{\partial}^{H}\mathbf{\widetilde{H}}_{0}(\bm{\phi})}{\partial{\theta}_{0}}\mathbf{C}^{-1} \frac{\partial\mathbf{\widetilde{H}}_{0}(\bm{\phi})}{\partial{\theta}_{0}}\mathbf{x}\},\\
        {F}_{{\theta}_{0},{\theta}_{\rm RIS}}&=2{\xi}_{0}^{2}\Re\{\mathbf{x}^{H}\frac{{\partial}^{H}\mathbf{\widetilde{H}}_{0}(\bm{\phi})}{\partial{\theta}_{0}}\mathbf{C}^{-1} \frac{\partial\mathbf{\widetilde{H}}_{0}(\bm{\phi})}{\partial{\theta}_{\rm RIS}}\mathbf{x}\},\\
        {F}_{{\theta}_{\rm RIS},{\theta}_{0}}&=2{\xi}_{0}^{2}\Re\{\mathbf{x}^{H}\frac{{\partial}^{H}\mathbf{\widetilde{H}}_{0}(\bm{\phi})}{\partial{\theta}_{\rm RIS}}\mathbf{C}^{-1} \frac{\partial\mathbf{\widetilde{H}}_{0}(\bm{\phi})}{\partial{\theta}_{0}}\mathbf{x}\},\\
        {F}_{{\theta}_{\rm RIS},{\theta}_{\rm RIS}}&=2{\xi}_{0}^{2}\Re\{\mathbf{x}^{H}\frac{{\partial}^{H}\mathbf{\widetilde{H}}_{0}(\bm{\phi})}{\partial{\theta}_{\rm RIS}}\mathbf{C}^{-1} \frac{\partial\mathbf{\widetilde{H}}_{0}(\bm{\phi})}{\partial{\theta}_{0}}\mathbf{x}\},\\
        \mathbf{F}_{{\theta}_{0},\bm{\alpha}_{0}^{T}}&=2\Re\{{\alpha}_{0}^{*}\mathbf{x}^{H}\frac{{\partial}^{H}\mathbf{\widetilde{H}}_{0}(\bm{\phi})}{\partial{\theta}_{0}}\mathbf{C}^{-1}[1~\jmath]^T\otimes\mathbf{\widetilde{H}}_{0}(\bm{\phi})\mathbf{x}\},\\
        \mathbf{F}_{{\theta}_{\rm RIS},\bm{\alpha}_{0}^{T}}&=2\Re\{{\alpha}_{0}^{*}\mathbf{x}^{H}\frac{{\partial}^{H}\mathbf{\widetilde{H}}_{0}(\bm{\phi})}{\partial{\theta}_{\rm RIS}}\mathbf{C}^{-1}[1~\jmath]^T\otimes\mathbf{\widetilde{H}}_{0}(\bm{\phi})\mathbf{x}\},\\
        \mathbf{F}_{\bm{\alpha}_{0}\bm{\alpha}_{0}^{T}}& = 2\Re\big\{\{[1~\jmath]^T\otimes \mathbf{\widetilde{H}}_{0}(\bm{\phi})\mathbf{x}\}^H\mathbf{C}^{-1}[1~\jmath]^T\otimes\mathbf{\widetilde{H}}_{0}(\bm{\phi})\mathbf{x}\big\}.
    \end{align}
\end{subequations}
Then, after having $\mathbf{F}_{\bm{\alpha}_{0}\bm{\alpha}_{0}^{T}}$, the sub-matrices  $\mathbf{F}_{\bm{\theta}\bm{\theta}^T}$ and $\mathbf{F}_{\bm{\theta}\bm{\alpha}_{0}^T}$ can be constructed as
\begin{align}   \mathbf{F}_{\bm{\theta}\bm{\theta}^T}= \begin{bmatrix}
    {F}_{{\theta}_{0},{\theta}_{0}}  & {F}_{{\theta}_{0},{\theta}_{\rm RIS}}\\
    {F}_{{\theta}_{\rm RIS},{\theta}_{0}}  & {F}_{{\theta}_{\rm RIS},{\theta}_{\rm RIS}}
    \end{bmatrix},
    \mathbf{F}_{\bm{\theta}\bm{\alpha}_0^T} = \begin{bmatrix}
    \mathbf{F}_{{\theta}_{0},\bm{\alpha}_0^T}\\
    \mathbf{F}_{{\theta}_{\rm RIS},\bm{\alpha}_{0}^T}
    \end{bmatrix}.
\end{align}

\end{document}